# Fundamentals of Caching Layered Data objects

AGRIM BARI, The University of Texas at Austin, USA
GUSTAVO DE VECIANA, The University of Texas at Austin, USA
GEORGE KESIDIS, The Pennsylvania State University, USA

The effective management of large amounts of data processed or required by today's cloud or edge computing systems remain a fundamental challenge. This paper focuses on cache management for applications where data objects can be stored in layered representations. In such representations, each additional data layer enhances the 'quality' of the object's version but comes with an incremental cost of memory space. This layered approach proves beneficial in various scenarios, including the delivery of zoomable maps, video coding, future Virtual Reality gaming, and layered neural network models where additional data layers improve inference accuracy. In systems where users or devices demand different versions of a data object, layered representations offer flexibility for caching policies to achieve improved hit rates.

In this paper, we explore the performance of various traditionally studied caching policies, such as Belady, LRU, and LFU, both with and without layering. To this end, we develop an asymptotically accurate analytical model for Layered LRU (LLRU). We study how the performance of LLRU is impacted by factors such as the number of layers, the popularity of different objects and layers, and overheads associated with storing layered representations. For instance, we show that, for LLRU, more layers are not always beneficial and indeed performance depends in subtle ways on the popularity and size profiles of layers.

## 1 INTRODUCTION

***Managing shared edge caching.*** Efficient management of shared memory systems for applications requiring large amounts of data continues to be a challenging problem. These challenges are exacerbated when mobile applications with latency constraints leverage limited/costly edge caching resources but have limited or variable connectivity to the network edge. In such settings, ensuring data is available when needed is all the more critical.

***Layered representations and applications.*** The focus of this paper is on applications where data objects can be stored and be of use in multiple versions which are encoded in Layered Representations (LRs). Each version of a data object embodies a tradeoff in size, and thus resource requirements, versus the 'quality' that an application can extract. LRs are such that the cumulative availability of each additional layer delivers a version with improved quality. Such an incremental approach to representing data object versions brings flexibility to systems where applications have heterogeneous quality requirements or can tolerate quality degradation when resources are scarce. LRs have found applications in, e.g., zoomable maps, video compression, and Virtual Reality (VR) games. For maps, LRs can be used to deliver different levels of topographic detail. Similarly, scalable video coding includes a base layer that contains essential information for lower quality, while enhancement layers contribute details for higher quality. In computer graphics, particularly relevant to VR, progressive meshes [13] can be used for efficient storage and rendering of 3D models. Through iterative mesh simplification algorithms, along with vertex splitting and collapse operations, a hierarchical structure emerges. Each level in this hierarchy represents a progressively more detailed version of the original mesh. LRs are also applicable when seeking to have compact Neural Network (NN) models. In this context, the base NN model can deliver lower inference accuracy but can be enhanced through additional data layers, which increase model complexity or weight fidelity, resulting in an NN model with higher inference accuracy [17, 18].

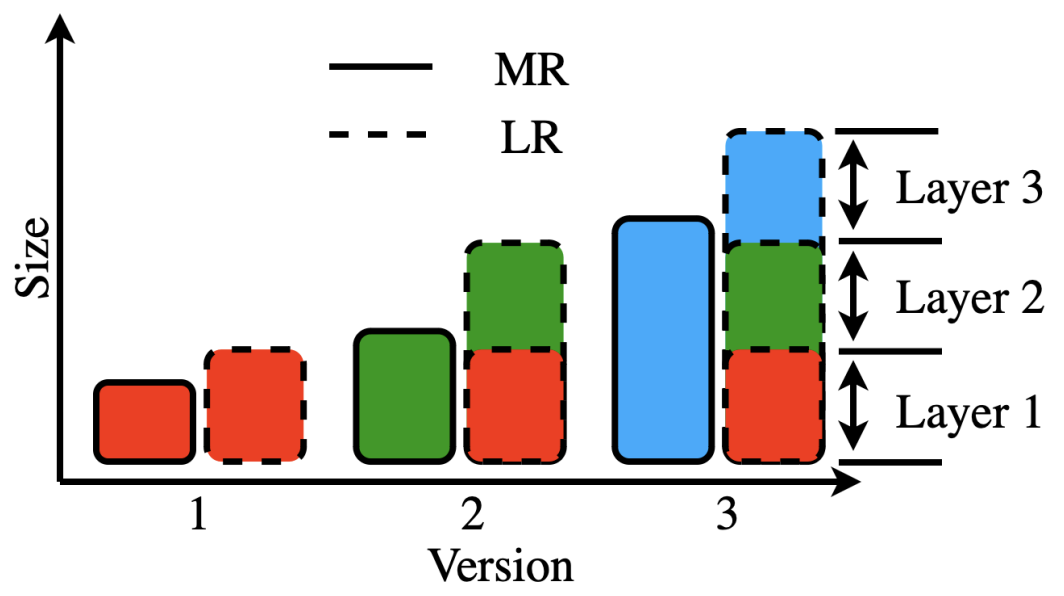

Fig. 1. 3 Versions under Multiple/Layered representation of a data object.

***Exploiting layered representations.*** Applications may request different versions of a data object for various reasons. First, an application might consider the computational resources of the end device, e.g., an end device with limited resources may request a version that requires less compute/memory/energy. Second, data object requests may vary based on the communication network conditions. In instances where the end device has limited bandwidth, preventing it from receiving high-quality data promptly, users may opt for versions that balance quality with efficient transmission. Thirdly, an application may simply not require the highest quality version. For example, in VR gaming settings, a detailed model for a complex tree that is far away would not be visible and thus is not required. Overall, these diverse considerations highlight the needed flexibility that LRs would be able to satisfy for a range of application requirements and constraints.

***Alternative representations.*** There are other ways of representing different versions of data objects, e.g., Multiple Representations (MRs) with or without transcoding, see [11, 21, 23]. In the case of MR without transcoding, discrete and independent versions of data objects are created — in the context of video streaming, these versions correspond to encoding video at different rates without any layering. MR may be more storage-efficient as compared to LR for the same version because LR may require additional information to extend to higher versions, see Figure 1. However, LR may be more storage efficient when plural quality/resolution levels of the same object are simultaneously in demand because the MR version will have a significant amount of identical (lower quality) information.

Whereas, MR with transcoding involves storing only one version corresponding to the highest 'quality' level - so, any lower version can be readily computed from this version. This transcoding can occur either in real-time (online transcoding), generating lower versions in real-time upon request, or in non-real-time (offline transcoding) where transcoding takes place in the backend for potential future requests. However, in this paper, we will focus solely on MR without transcoding limiting the computational burdens and delays associated with real-time transcoding.

***Caching policies with layered representations.*** While considerable attention has been paid to the design and analysis of caching policies for MR, limited attention has been given to the LR setting which as mentioned above, we expect will be of increasing relevance to emerging applications and caching at the network edge. In this paper, we focus on a disciplined study of traditional caching policies which have been redesigned to leverage LR data representations. Below we briefly discuss relevant related research before summarizing our contributions.

## 1.1 Related work

***Analytical works on caching policies and approximation.*** We restrict our review to papers most relevant to our work. [5, 15] summarize significant early work in the design of caching policies, and [12] describes analytical methods and evaluation results for the performance assessment of caching strategies. The aim of any caching policy is to achieve efficient cache utilization. This efficiency is measured primarily by the cache hit rate, which is the averaged fraction of data object requests for which the data object is in the cache when requested.

Besides hit rate, other design objectives for caching policies are ease of implementation, low operational overhead, and adaptability to fluctuations in access/request patterns. An important difference among caching policies is in what they evict when the cache is full. Under Least Recently Used (LRU), the cache is consistently updated to hold the most recently requested data objects, enabling it to leverage the temporal locality of data object requests. Notably for LRU under the Independent Reference Model (IRM), where each data object is requested independently of any past requests, the invariant distribution assuming data objects of the same size [19] and an approximation for the hit rate [4, 6, 7, 10] have been obtained. In particular, [6] describes the working-set approximation for hitting probabilities, the fraction of requests for a data object for which the object is in the cache. This approximation has been shown to be accurate as the number of objects scales [7, 10]. We herein extend the analysis of [6] for the approximation to our setting, where data objects have layered representations, and also demonstrate the asymptotic accuracy of the approximation based on ideas from [7].

Under the IRM model, for a fixed cache capacity with same-size data objects, caching the most popular data objects is optimal for causal policies [1]. Least Frequently Used (LFU) performs optimally under stationary regimes of request patterns by replacing cached data objects based on the frequency measurements of past requests. An interesting work by [14] shows that a variant of LRU that infers the instantaneous request rate subject to the history of requests can come arbitrarily close to the optimal LFU algorithm. [16] shows that even for strongly correlated request patterns, LFU is still optimal among causal policies. However, while LFU may be effective in stationary scenarios where access patterns remain relatively constant, it may struggle to perform optimally in non-stationary regimes where the dynamics of data access change over time.

Moving beyond the IRM, various researchers have conducted competitive analyses, considering the total number of cache misses as the figure of merit [9, 24]. They compare the performance of an online policy, i.e., one that makes eviction decisions without knowledge of future requests, with the optimal offline policy, Belady [3], i.e., one that knows the entire sequence of requests in advance. The LRU policy has a competitive ratio that scales linearly with the cache size, $B$. Improving on LRU, researchers in [9] have shown that a randomized online algorithm, Marker, which uses markers to decide and prioritize critical data objects, could be worse than the optimal offline algorithm by a factor of $2H_B \approx 2\log(B)$ ($H_B$ denotes the $B$th harmonic number: $H_B = 1 + 1/2 + 1/3 + \ldots + 1/B$), but not more. Moreover, no online algorithm could achieve a factor less than $H_B$. Recently, [20] extended these results to cases where the traditional marker algorithm is combined with predictions about the next time of request for objects currently in the cache when making eviction decisions.

If done correctly, they show that one can improve upon this factor of $\log(B)$ depending on the accuracy of the prediction.

***Multiple vs. Layered Representations.*** Researchers have also explored the caching problem for objects which could either be in MR or LR. In [11], the authors advocate for storing MR for some data objects and LR for the rest when the goal is to maximize the hit rate. In addition, they also develop heuristic policies that dynamically adapt the representation for each data object. [22] compares optimization-based static caching policies for MR versus LR to conclude that the hit rate for an LR based caching policy is superior. We address this question more broadly by showing the benefit of LR over MR in terms of hit rate as a function of MR's storage efficiency compared to LR. We also present results showing the benefit of LR as we vary the cache size, the fraction of requests for different versions, the relative size of versions, and the number of versions.

## 1.2 Paper contributions and organization

The main contributions of this paper are now summarized. First, we redesign and analyze traditional caching policies (Belady, LFU, LRU) for settings where data objects are available in MR or LR. In particular, we introduce a new working set approximation to compute the hit probabilities for data objects in a cache utilizing Layered LRU (LLRU) caching policy under an IRM for requests for data object versions. We show the asymptotic accuracy of this approximation for both a fixed number of layers and a continuum of layered representations. The continuum model seems appropriate for settings where layering overhead is minimal and thus applications could in principle cache only the data it requires for where it needs better quality, e.g., in a VR gaming setting where high quality is needed only for aspects of the environment that are currently (or maybe in the near future) close by.

Second, using the working-set approximation, we evaluate the benefit of LR versus MR for a fixed set of equivalent data object versions. Our results suggest that even if LR incurs relatively high overheads versus MR, the performance benefits of LR representations are excellent. We note however that this does depend on the popularity of the distinct versions, i.e., the layered structure is particularly beneficial when there is sufficient diversity in the requests for a data object's versions. With these observations in mind, we consider greedy caching policies that might exploit the availability of *both* LR and MR, by greedily seeking to represent the versions in the cache in the most memory-efficient manner. Such policies can provide some benefit but only under highly skewed popularity for data object versions. We also explore the performance of various layered caching policies under stationary IRM showing that layered LRU is not quite on par either with layered LFU or, of course, the genie-based layered Belady; yet LLRU can be expected to be a workhorse for caching LR based systems because of its simplicity and robustness to dynamic request distribution.

Finally, we explore the performance sensitivity of LLRU to the size and popularity of layers and data object versions. This provides an avenue to study how many layers are enough or when indeed more layers leads to better performance.

The paper is organized as follows. We start by describing the system model and working-set approximation for the LLRU policy in Section 2. In the same section, we describe the re-design of traditional caching policies (Belady, LFU, LRU) with LRs along with an optimization-based static-offline caching policy. In Section 3 we empirically evaluate the claims of Section 2. Finally, Section 4 concludes the paper.

## 2 SYSTEM MODEL AND ANALYSIS

### 2.1 Model for cache

The system consists of a cache server of capacity $B$ bytes. The server stores various versions of data objects to serve near-future requests from a user population. Owing to practical constraints, the cache capacity typically is not enough to store all versions of data objects.

### 2.2 Model for data objects and arrival requests

We let $\mathcal{D}$ denote the set of these data objects - the set has cardinality $D = |\mathcal{D}|$. Each data object $d \in \mathcal{D}$ can be stored in several versions, $v \in \{1, 2, \dots, V\}$, where $V$ is the number of versions. We adopt the Independent Reference Model (IRM), which is a good abstraction for independent requests generated from a large population of users. Let $\lambda(d, v)$ denote the arrival rate of requests for version $v$ of data object $d$. The total arrival rate of requests for data object $d \in \mathcal{D}$ is given by $\lambda(d) = \sum_{v=1}^{V} \lambda(d, v)$ and $\lambda = \sum_{d=1}^{D} \lambda(d)$ denotes the total arrival rate of requests generated by a population of users. We denote the vector of requests for each version and data object as $\boldsymbol{\lambda} = (\lambda(d, v) : d \in \mathcal{D}, v \in \{1, 2, \dots, V\})$. We define $q(d) = \lambda(d)/\lambda$ and $q(d, v) = \lambda(d, v)/\lambda$ as the probability of request for data object $d \in \mathcal{D}$ and probability of request for data object $d \in \mathcal{D}$ in version $v$, respectively. We consider two representations for storing these data object versions as explained below.

### 2.3 Model for Multiple Representations

Under *Multiple Representations* (MRs), several distinct versions of a data object can be maintained in the cache. Let $s_{\text{MR}}(d, v)$ denote the cache storage space occupied by data object $d \in \mathcal{D}$ in version $v$ under MR. The size of versions of a data object $d$ under MR is strictly increasing in $v$, i.e., $s_{\text{MR}}(d, 1) < s_{\text{MR}}(d, 2) < \cdots < s_{\text{MR}}(d, V)$. As explained before, if one version of a data object is cached and there is a request for a different version of the same data object, the cached version cannot be used to serve this request under multiple representations.

### 2.4 Model for Layered representations

Under *Layered Representations* (LRs), a version $v$ of a data object is represented by a set of consecutive layers $l \in \{1, 2, \dots, v\}$ where the size of layer $l$ for data object $d$ is denoted by $\delta(d, l)$. So version $v$ of data object $d$ occupies $s_{\text{LR}}(d, v) = \sum_{l=1}^{v} \delta(d, l)$ space in the cache. Note that the incremental layer sizes $\delta(d, l)$ need not be strictly increasing or decreasing in $l$. We will be exploring the impact of this in later sections. Depending on the application, we expect the overall size of representations under LR to be larger than MR, i.e., $s_{\text{MR}}(d, v) \leq s_{\text{LR}}(d, v)$. We will be studying the impact of such overheads in the sequel. We let $\gamma(d, l) = \sum_{v=l}^{V} \lambda(d, v)$ denote the total request rate for layer $l$ induced by the requests for different versions of data object $d$ and $p(d, l) = \gamma(d, l)/\lambda = \sum_{v=l}^{V} q(d, v)$ denote the request probability for layer $l$ of data object $d$.

### 2.5 Caching Policies

We consider a set $\Pi$ of caching policies. These policies can either be online or offline, i.e., they adapt the cached content based on incoming requests or not, respectively, and they may have knowledge about the future requests or request rate. For a given vector of request rate $\boldsymbol{\lambda}$ and policy $\pi \in \Pi$, we define $\mathbf{h}_{\boldsymbol{\lambda},\pi} = (h_{\boldsymbol{\lambda},\pi}(d, r) : d \in \mathcal{D}, v \in \{1, 2, \dots, V\})$, where $h_{\boldsymbol{\lambda},\pi}(d, v)$ denotes the long-term fraction of requests for data object $d$ and version $v$ that results in a cache hit. These data objects could either be stored in LR or MR.

### 2.6 Performance metric

We capture the overall performance of the cache in terms of hit rate. For a given vector of request rates $\boldsymbol{\lambda}$ under policy $\pi$, we define it as

$$H_{\boldsymbol{\lambda},\pi} = \sum_{d=1}^{D}\sum_{v=1}^{V} \lambda(d,v) h_{\boldsymbol{\lambda},\pi}(d,v). \tag{1}$$

### 2.7 Layered Caching Policies

We now introduce our caching policies for layered representations. Note that for all policies, a user request to access an object involves using a hash table to determine whether the object is cached (i.e., whether it's a cache hit) and, if so, where it is stored in cache memory. We first define a common property of all layered caching policies stated hereafter.

***Property of layered caching policies.*** For all policies discussed hereafter, if layer $l+1$ is present in the cache, then layer $i \in \{1,2,\ldots,l\}$ is also present in the cache.

*2.7.1 Static optimal.* We begin by developing an optimization-based static-caching policy that maximizes the hit rate *given* the vector of request rates $\boldsymbol{\lambda}$, where data objects are in LRs. This is the best that a policy with no knowledge of future requests can do. Let $\mathbf{x} = (x(d,v) : d \in \mathcal{D}, v \in \{1,2,\ldots,V\})$, where $x(d,v)$ denotes an indicator for whether data object $d$ in version $v$ is included in the cache or not. We formulate the following optimization that maximizes the hit rate.

$$\max_{\mathbf{x}} \quad \sum_{d=1}^{D}\sum_{v=1}^{V} \lambda(d,v)x(d,v) \tag{2a}$$

$$\text{s.t.} \quad \sum_{d=1}^{D}\sum_{v=1}^{V} \delta(d,v)x(d,v) \le B, \tag{2b}$$

$$x(d,v-1) \ge x(d,v), \quad \forall\, d \in \mathcal{D}, v \in \{2,3,\ldots,V\}, \tag{2c}$$

$$x(d,v) \in \{0,1\} \qquad \forall\, d \in \mathcal{D}, v \in \{1,2,\ldots,V\} \tag{2d}$$

where Constraint 2b is on the cache capacity and Constraint 2c is ensuring the previously mentioned property of layered caching policies.

*2.7.2 Layered Least Frequently Used (LLFU).* The LLFU caching policy prioritizes the caching of layers of data objects that have been accessed most frequently while ensuring Constraint 2c. Although we will show through simulations that this policy is optimal when the size of each layer is equal among policies with no knowledge of future requests, it involves tracking and updating access frequencies of each layer of every data object. Thus, it may not be practical. An LLFU cache serves an incoming request for object $(d,v)$ as follows:

- The number of accesses for layer $(d,l)$ is incremented for all $l \le v$.
- If a cache miss occurs (i.e., layer $(d,v)$ is not present in the cache), to meet Constraint 2b, the server may need to evict layers of cached data objects in increasing order of their current number of accesses until there is enough space to store all layers $l \le v$ for data object $d$.

So, under LLFU, only layers with the currently highest access counts are cached. Periodically to prevent numerical overflow, access counts *of all data object layers* can, e.g., be simultaneously decremented by a common amount (equal to the currently smallest access count among all data object layers). Instead of access counts, one can define an LLFU policy with access frequencies equal to the inverse of auto-regressive estimates of inter-access times.

A hybrid LR-MR LFU policy is described in Section 2.10 below.

*2.7.3 Layered Least Recently Used (LLRU).* LLRU manages the cache by evicting the least recently accessed layers among all data objects currently in the cache. This policy works well when there is temporal locality of request patterns, i.e., layers and data objects accessed more recently are more likely to be accessed again in the near future. Let $a(d, l)$ denote the time of last access for layer $l$ and data object $d \in \mathcal{D}$. An LLRU cache serves an incoming request for $(d, v)$ at time $t$ as follows:

- Set $a(d, l) = t$ for all $l \leq v$ while ensuring that lower layers come after the higher layers.
- If a cache miss occurs, to meet Constraint 2b the server may need to evict layers of cached data objects in the increasing order of access times until there is enough space to store all layers $l \leq v$ for data object $d$.

Instead of using access times, the LLRU cache-eviction order can be maintained by just using a doubly-linked list.

A hybrid LR-MR LRU policy is described in Section 2.10 below.

*2.7.4 Layered Belady (LBelady).* LBelady evicts by identifying layers of data objects that will be accessed furthest in the future and is thus a non-causal policy. Let $f(d, l; t) > t$ denote the smallest access time after $t$ for layer $l$ of data object $d$. When the size of each layer is equal, this is the optimal policy among all possible policies, albeit accurate future knowledge is generally not available. In our simulations, this will serve as a benchmark for the case of equal layer sizes. An LBelady cache serves an incoming request at time $t$ in the following manner:

- If a cache miss occurs, to meet Constraint 2b, the server may evict layers of data objects $(d, l)$ in decreasing order of $f(d, l; t)$ until there is enough space to store the requested version.

## 2.8 Working-set approximation for LLRU

We present a working-set approximation for the LLRU policy when data objects are stored in layered representations. We will demonstrate its accuracy by studying this working set as number of data objects go to infinity in the next section and in a later section through simulations.

Consider a system where time is divided into slots. For the analysis we assume that the request arrival process for each data object $d$ and layer $l$ is a Bernoulli process with parameter $p(d, l)$, i.e., the probability that there is a request for data object $d$ and layer $l$ in a time slot is $p(d, l)$ and such events occur independently across time slots.

*2.8.1 Characteristic time.* Suppose there is a request for data object $d$ and layer $l$ at time zero. Let $T_f(i, k)$ be the time of first request for data object $i \neq d$ and layer $k$, where $k = \{1, 2, \ldots, V\}$. We use $T_n(d, m)$ to denote the time of next query for data object $d$ and layer $m$, where $m \leq l$. Under the Bernoulli arrival process model, these times are geometrically distributed, i.e., $T_f(d, l) \sim \text{Geo}(p(d, l))$ or $T_n(d, l) \sim \text{Geo}(p(d, l))$.

At time $t > 0$, the total size of different data objects and layer requested up to time $t$ (i.e., working-set size), excluding requests for data object $d$ and layer $l$ is given by:

$$S_{-(d,l)}(t) = \sum_{\substack{i=1 \\ i \neq d}}^{D} \sum_{k=1}^{V} \delta(i, k) \mathbf{1} \left\{ T_f(i, k) < t \right\} + \sum_{k=1}^{l-1} \delta(d, k) \mathbf{1} \left\{ T_n(d, k) < t \right\}, \tag{3}$$

where $\delta(i, k)$ represents the size of layer $k$ for data object $i$.

The characteristic time $T_{-(d,l)}(B)$, a random variable, is defined as the minimum time $t > 0$ at which the working-set size excluding data object $d$ and layer $l$ exceeds $B$:

$$T_{-(d,l)}(B) = \min\{t > 0 : S_{-(d,l)}(t) \geq B\}. \tag{4}$$

A request for data object $d$ and layer $l$ at time $T_n(d,l)$ is a cache hit if the working-set size remains below $B$, i.e., $S_{-(d,l)}(T_n(d,l)) < B$, or equivalently, if $T_n(d,l) < T_{-(d,l)}(B)$. This relationship is expressed as:

$$\{S_{-(d,l)}(T_n(d,l)) < B\} = \{T_{-(d,l)}(B) > T_n(d,l)\}. \tag{5}$$

Thus, the hit probability for data object $d$ and layer $l$ is then

$$h(d,l) = \mathbb{P}\left(T_{-(d,l)}(B) > T_n(d,l)\right) = \mathbb{E}\left[1 - (1-p(d,l))^{\left(T_{-(d,l)}(B)-1\right)}\right]. \tag{6}$$

Since $T_{-(d,l)}(B)$ corresponds to the time when the working-set size first reaches $B$, we have:

$$B = \sum_{\substack{i=1\\i\neq d}}^{D}\sum_{k=1}^{V}\delta(i,k)\mathbf{1}\left\{T_f(i,k) < t\right\} + \sum_{k=1}^{l-1}\delta(d,k)\mathbf{1}\left\{T_n(d,k) < t\right\}. \tag{7}$$

and taking expectations on both sides and simplifying,

$$B = \sum_{\substack{i=1\\i\neq d}}^{D}\sum_{k=1}^{V}\delta(i,k)\mathbb{E}\left[1-(1-p(i,k))^{T_{-(d,l)}(B)-1}\right] + \sum_{k=1}^{l-1}\delta(d,k)\mathbb{E}\left[1-(1-p(d,k))^{T_{-(d,l)}(B)-1}\right]. \tag{8}$$

We use two common approximations from the literature to simplify hit probability calculations; see [4, 10] for details.

Approximation 1: For $D \gg 1$, the characteristic time $T_{-(d,l)}(B)$ becomes concentrated around its mean value. Therefore, $T_{-(d,l)}(B)$ can be approximated by a deterministic value $t_{-(d,l)}(B)$ for data object $d$ and layer $l$. Thus, the above equation can be rewritten as follows:

$$B = \sum_{\substack{i=1\\i\neq d}}^{D}\sum_{k=1}^{V}\delta(i,k)\left(1-(1-p(i,k))^{t_{-(d,l)}(B)-1}\right) + \sum_{k=1}^{l-1}\delta(d,k)\left(1-(1-p(d,k))^{t_{-(d,l)}(B)-1}\right). \tag{9}$$

The above is a fixed point equation, which can solved to find $t_{-(d,l)}(B)$ and one can use that to find the hit probability for data object $d$ and layer $l$ by

$$h(d,l) = \left(1-(1-p(d,l))^{\left(t_{-(d,l)}(B)-1\right)}\right). \tag{10}$$

Approximation 2: The dependence of $t_{-(d,l)}(B)$ on $(d,l)$ can be ignored for all data objects and layer. This is works when $p(d,l)$ is relatively insignificant to 1, and becomes exact if request probabilities are equiprobable. In summary, the working-set approximation for LLRU is as follows. Let $t^*(B)$ be such that:

$$B = \sum_{d=1}^{D}\sum_{l=1}^{V}\delta(d,l)\left(1-(1-p(d,l))^{(t^*(B)-1)}\right) \tag{11}$$

Then the hit probability for data object $d \in \mathcal{D}$ and layer $l \in \{1,2,\ldots,V\}$ is given by

$$h(d,l) = \left(1-(1-p(d,l))^{(t^*(B)-1)}\right). \tag{12}$$

This hit probability for data object $d$ and layer $l$ is equal to hit probability for data object $d$ and version $v$, where $v = l$ because of the property of LLRU. The results for a time-slotted system can be extended to continuous time, where the request arrival process for data object $d$ and layer $l$ is a Poisson process with parameter $\gamma_{d,l}$. The hit probability for data object $d$ and layer $l$ is given by

$$h(d,l) = 1 - e^{-\gamma_{d,l}t^*(B)}, \tag{13}$$

where $t^*(B)$ is such that:

$$B = \sum_{d=1}^{D} \sum_{l=1}^{V} \delta(d,l) \left(1 - e^{-\gamma_{d,l} t^*(B)}\right). \tag{14}$$

In the next section, we show the asymptotic accuracy of working-set approximation.

### 2.9 Asymptotic accuracy of working-set approximation for LLRU

We extend the analysis from [7] to incorporate layers into the construction, focusing on LR for this part. Specifically, we consider a system of caches where the request probability for data objects and the working-set size scale as a function of $D$. In this framework, each data object is assumed to have $V$ fixed layers (or versions).

Let $F$ be a smooth, monotone increasing function with domain $[0, 1]$, such that $F(0) = 0$ and $F(1) = 1$. We define the request probability for data object $d$ and version $v$ as $D$ scales in the following manner:

$$q^{(D)}(d,v) = (F(d/D) - F((d-1)/D))\, g(v; d/D) \tag{15}$$

where $g(v; d/D)$ denotes the request probability for version $v$ of data object $d$ and $\sum_{v=1}^{V} g(v; d/D) = 1$ for all data objects. Based on the definition of $F$ and $g$, we have $\sum_{d=1}^{D} \sum_{l=1}^{V} q^{(D)}(d,v) = 1$ and $q^{(D)}(d,v) \geq 0$. Thus, $q^{(D)}(d,v)$ is a probability distribution determined by $F$ and $g$. We use $\delta^{(D)}(d,l)$ to denote the size of layer $l$ for data object $d$ and $p^{(D)}(d,l) = \sum_{v=l}^{V} q^{(D)}(d,v)$ denotes the request probability for layer $l$ of data object $d$.

We define $b = B/D$, which scales as a function of $D$, and develop the notion of characteristic time in the same way as in the previous section. We assume a system with time-slots and request arrival process for data object $d$ and layer $l$ is a Bernoulli process with parameter $p^{(D)}(d,l)$ and such events occur independently over time-slots.

Suppose there is a request for data object $d$ and layer $l$ at time zero. Let $T_f^{(D)}(i,k)$ be the time of first request for data object $i \neq d$ and layer $k$, where $k = \{1, 2, \ldots, V\}$. We use $T_n^{(D)}(d,m)$ to denote the time of next query for data object $d$ and layer $m$, where $m \leq l$. Under the Bernoulli arrival process model, these times are geometrically distributed, i.e., $T_f^{(D)}(d,l) \sim \text{Geo}(p^{(D)}(d,l))$ or $T_n^{(D)}(d,l) \sim \text{Geo}(p^{(D)}(d,l))$. At time $t > 0$, the total size of different data objects and layer requested upto time $t$ (i.e., working-set size), excluding requests for data object $d$ and layer $l$ is

$$S_{-(d,l)}^{(D)}(t) = \sum_{k=1}^{l-1} \delta^{(D)}(d,k) \mathbf{1}\left\{T_n^{(D)}(d,k) < t\right\} + \sum_{\substack{i=1 \\ i\neq d}}^{D} \sum_{k=1}^{V} \delta^{(D)}(i,k) \mathbf{1}\left\{T_f^{(D)}(i,k) < t\right\}, \tag{16}$$

with

$$\mathbb{E}\left[S_{-(d,l)}^{(D)}(t)\right] = \sum_{k=1}^{l-1} \delta^{(D)}(d,k) \left(1 - (1 - p^{(D)}(d,k))^{(t-1)}\right) + \sum_{\substack{i=1 \\ i\neq d}}^{D} \sum_{k=1}^{V} \delta^{(D)}(i,k) \left(1 - (1 - p^{(D)}(i,k))^{(t-1)}\right). \tag{17}$$

Similarly, we can find the working-set size at time $t$ and its expectation is given by:

$$\mathbb{E}\left[S^{(D)}(t)\right] = \sum_{d=1}^{D} \sum_{l=1}^{V} \delta^{(D)}(d,l) \left(1 - (1 - p^{(D)}(d,l))^{(t-1)}\right). \tag{18}$$

We define Riemann integrable $\Delta$ satisfying $\Delta(d/D, l) = \delta^{(D)}(d,l)$ for all $D$, $d$ and $l$ for the theorem below.

Theorem 1 (Asymptotic hit probability). *Consider the system of caches which scales as a function of $D$. For large $D$, the hit probability for data object $d$ and layer $l$, $h^{(D)}(d,l)$, is approximated by*

$$h^{(D)}(d,l) = \left(1 - (1 - p^{(D)}(d,l))^{(t^*(B)-1)}\right) \tag{19}$$

*where $t^*(B)$ is such that:*

$$B = \sum_{d=1}^{D}\sum_{l=1}^{V} \delta^{(D)}(d,l)\left(1 - (1 - p^{(D)}(d,l))^{(t^*(B)-1)}\right). \tag{20}$$

*In the limit $D \longrightarrow \infty$, the hit probability for data object $d$ and layer $l$ is given by:*

$$h(d,l) := \lim_{D\to\infty} h^{(D)}(d,l) = 1 - e^{-\tau^*(b)F'(d)\sum_{v=l}^{V} g(v;d)} \tag{21}$$

*where $\tau^*(b)$ is such that:*

$$b = \lim_{D\to\infty} \mathbb{E}\left[\frac{S^{(D)}(D\tau)}{D}\right] = \int_0^1 \sum_{l=1}^{V} \Delta(x,l)dx - \int_0^1 \sum_{l=1}^{V} \Delta(x,l)e^{-\tau^*(b)F'(x)\sum_{v=l}^{V} g(v;x)}dx. \tag{22}$$

Proof. Refer to the Appendix. □

We next study a system of caches where probability requests for data objects, layers, and working-set size scales as a function of $D$ and $V$.

As before, let $F$ and $G$ be two smooth, monotone increasing function with domain closed interval $[0, 1]$, such that $F(0) = G(0) = 0$ and $F(1) = G(1) = 1$. We define the request probability for data object $d$ and version $v$ as $D$ and $V$ scales as follows:

$$q^{(D,V)}(d,v) = (F(d/D) - F((d-1)/D))\,(G(v/V) - G((v-1)/V))\,. \tag{23}$$

Based on the definition of $F$ and $G$, we have $\sum_{d=1}^{D}\sum_{l=1}^{V} q^{(D,V)}(d,v) = 1$ and $q^{(D,V)}(d,v) \geq 0$. Thus, $q^{(D,V)}(d,v)$ is a probability distribution determined by $F$ and $G$. We use $\delta^{(D,V)}(d,l)$ to denote the size of layer $l$ for data object $d$ and $p^{(D,V)}(d,l) = \sum_{v=l}^{V} q^{(D,V)}(d,v)$ denotes the request probability for layer $l$ of data object $d$. We define Riemann integrable $\Delta$ satisfying $\Delta(d/D, l/V) = \delta^{(D,V)}(d,l)$ for all $D, V, d$ and $l$ for the theorem below.

Theorem 2. *Consider the system of caches which scales as a function of $D$ and $V$. For large $D$ and $V$, the hit probability for data object $d$ and layer $l$, $h^{(D,V)}(d,l)$, is approximated by*

$$h^{(D,V)}(d,l) = \left(1 - (1 - p^{(D,V)}(d,l))^{(t^*(B)-1)}\right) \tag{24}$$

*where $t^*(B)$ is such that:*

$$B = \sum_{d=1}^{D}\sum_{l=1}^{V} \delta^{(D,V)}(d,l)\left(1 - (1 - p^{(D,V)}(d,l))^{(t^*(B)-1)}\right). \tag{25}$$

*In the limit $D \longrightarrow \infty$ and $V \longrightarrow \infty$, the hit probability for data object $d$ and layer $l$ is given by:*

$$h(d,l) := \lim_{V\to\infty}\lim_{D\to\infty} h^{(D,V)}(d,l) = 1 - e^{-\tau^*(b)F'(d)G'(l)} \tag{26}$$

*where $\tau^*(b)$ is such that:*

$$b = \lim_{R\to\infty}\lim_{D\to\infty} \mathbb{E}\left[\frac{S^{(D,V)}(DV\tau)}{DV}\right] = \int_0^1\int_0^1 \Delta(x,y)dxdy - \int_0^1\int_0^1 \Delta(x,y)(e^{-\tau^*(b)F'(x)G'(y)})dxdy. \tag{27}$$

Proof. Similar to the proof of Theorem 1. □

### 2.10 Greedy hybrid LRU and LFU policies

In this subsection, we briefly describe two policies that adapt and store the best representation choice (MR or LR) for each cached data object's associated versions. These approaches are inspired by the work of [11] and aim to store objects as MR if there is a skewed popularity among its different versions. Conversely, if a data object is popular across multiple versions, it is stored as LR. Under our Greedy Hybrid LRU-type policy, when a request occurs for a version of a data object which is not present in the cache in any other version, the data object is fetched in the requested version and stored in its MR representation. If there is a request for a version of a data object which is different from a version that has already been cached, then both versions are stored as LR including all layers up to the maximum version requested. This policy is called "HLRU" in our numerical evaluation section.

Similarly we can define a static Greedy Hybrid LFU - type approach where the policy to decide which objects to include in the cache proceeds as follows. Data object versions are ranked in descending order of popularity none of which are initially cached. Take the most popular uncached object/version $(d, v)$: If no cached version of $d$ exists and there is room, then $(d, v)$ enters the cache in its MR representation. If another version of $d$ is in the cache and if there is room, then $(d, v)$ enters the cache as LR the other cached version of $d$ converts to LR. This process proceeds until the cache is full. This static Greedy Hybrid LFU policy is such that objects cached only in one version are stored as MR, otherwise LR. This policy is called "HLFU" in our numerical evaluation section. The described policies are greedy under the constraint that for all data object $d$ and version $v$, $s_{\text{MR}}(d, v) \leq s_{\text{LR}}(d, v)$ and, for all versions $v > 1$, $\min_{1 \leq u < v \leq V} s_{\text{MR}}(d, u) + s_{\text{MR}}(d, v) \geq s_{\text{LR}}(d, v)$.

## 3 NUMERICAL EVALUATION AND SIMULATION RESULTS

In this section, we perform extensive numerical evaluations based on the working set approximation and simulations of layered caching policies. The aim is to characterize the fundamental tradeoffs underlying the caching of data objects with LR and/or MR representations.

### 3.1 How accurate is the working set approximation for LLRU?

**Setting.** We consider a caching system with $D = 100$ data objects, each having $V = 4$ layered versions. The request probability $q(d)$ follows a Zipf distribution with parameter 0.8, while the request probability $q(d, v)$ for version $v$ is uniformly selected from $(0, q(d))$, ensuring $\sum_{v=1}^{V} q(d, v) = q(d)$. Additionally, we impose $q(d, v_1) > q(d, v_2)$ for $v_1 < v_2$, reflecting the higher request frequency of lower versions. Requests for object $d$ and version $v$ follow a Poisson process with rate $q(d, v)$. The size of each layer is uniformly chosen from $[1, 240]$, ensuring a total object size of 240.

**Results discussion.** We plot the hit probability of data objects ranked 1, 5, 10, and 15 in Figure 2. The squares represent the results obtained from simulations of LLRU policy, conducted over sufficiently long runs to ensure high accuracy. The lines are derived from the working set approximation for LLRU. The agreement between simulation results and approximation is nearly perfect for all practical purposes across all layers of a data object. We will use this approximation to address the questions posed at the beginning.

### 3.2 When are Multiple Representations (MR) better than Layered Representations (LR)?

**Setting.** We now examine a caching system with $D = 100$ data objects and $V = 2$ versions under multiple/layered representations, where request probability for a data object follows a Zipf distribution with parameter 0.8. Let $\alpha = q(d, 1)/(q(d, 1) + q(d, 2))$ denote the request probability for Version 1 of MR/LR for data object $d$. Thus, the request probability for Version 2 of either MR or LR is $1 - \alpha$. For the case of multiple representations, $\beta = s_{\mathrm{MR}}(d, 1)$ denotes the size of Version 1 and the size of Version 2 is 1, i.e., $s_{\mathrm{MR}}(d, 2) = 1$ for data object $d$. The size of Version 1 and 2 under layered representation is given by $s_{\mathrm{LR}}(d, v) = (1 + o) \cdot s_{\mathrm{MR}}(d, v)$, where $o$ is the percent overhead of LR vs. MR. As before the request arrival process is modeled as a Poisson process and we set the total request rate, $\lambda$, to = 1.

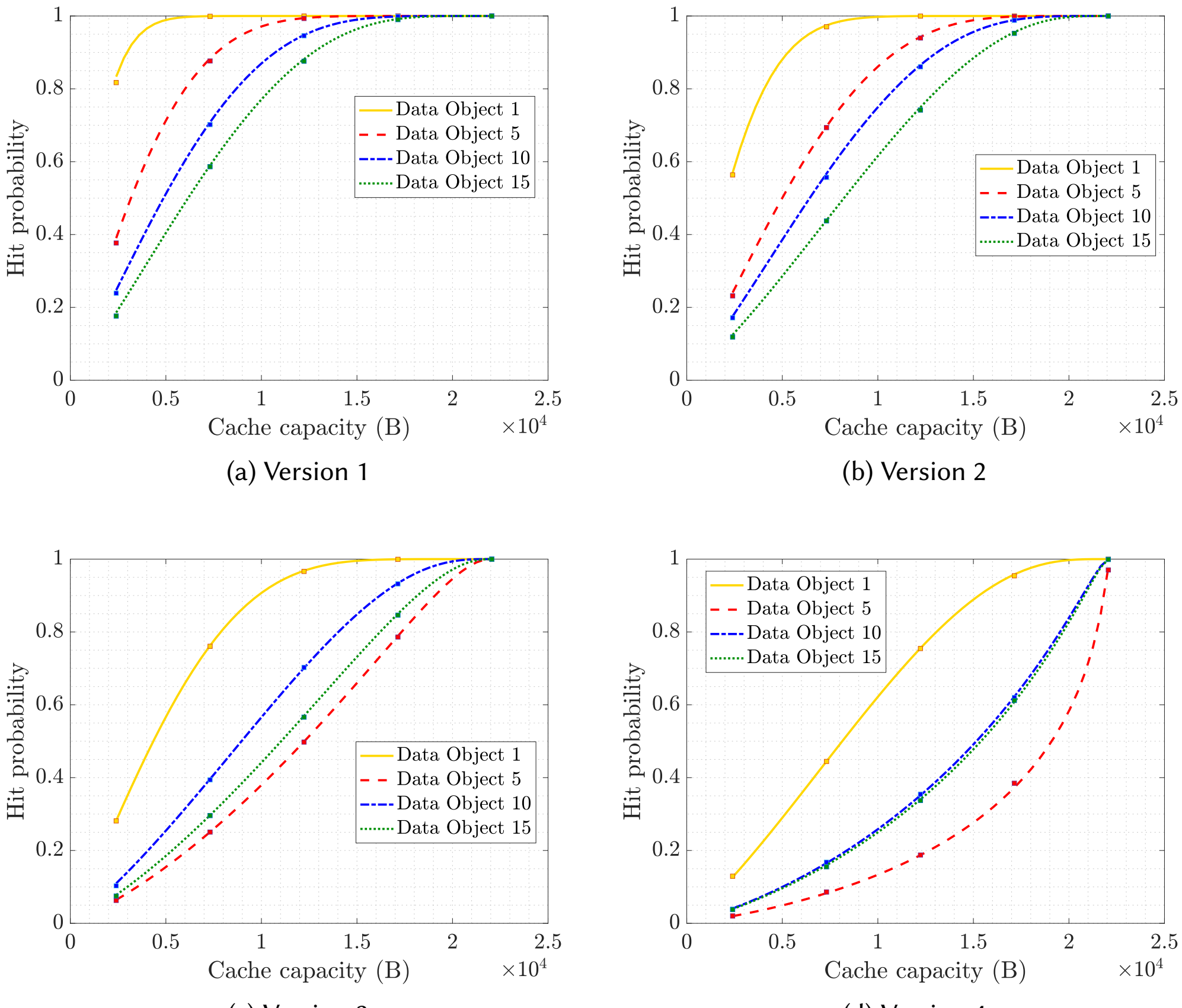

(a) Version 1

(b) Version 2

(c) Version 3

(d) Version 4

Fig. 2. Hit probability against cache capacity for selected data objects under LLRU caching policy.

**Results discussion.** In Figure 3a, we show the percentage relative improvement in the hit rate of LLRU (data objects are stored in LR) compared to MRLRU (data objects are stored in MR) for varying cache capacities (10, 20, and 100). As the cache capacity increases, the observed improvement decreases. This trend emerges because the hit rate for MRLRU and LLRU converges to 1 with increasing cache capacity, regardless of overhead. Ultimately, a sufficiently large cache achieves the optimal hit rate of 1. Consequently, for such large cache capacities, there will be no difference in hit rates between LLRU and MRLRU, leading to no relative improvement.

Moreover, in Figure 3b, we plot the hit rate under two scenarios: one where all data objects are exclusively stored in LR (with a cache utilizing LLRU) and another where they are stored in MR (with a cache utilizing MRLRU). This is presented as a function of fraction of requests for Version 1.

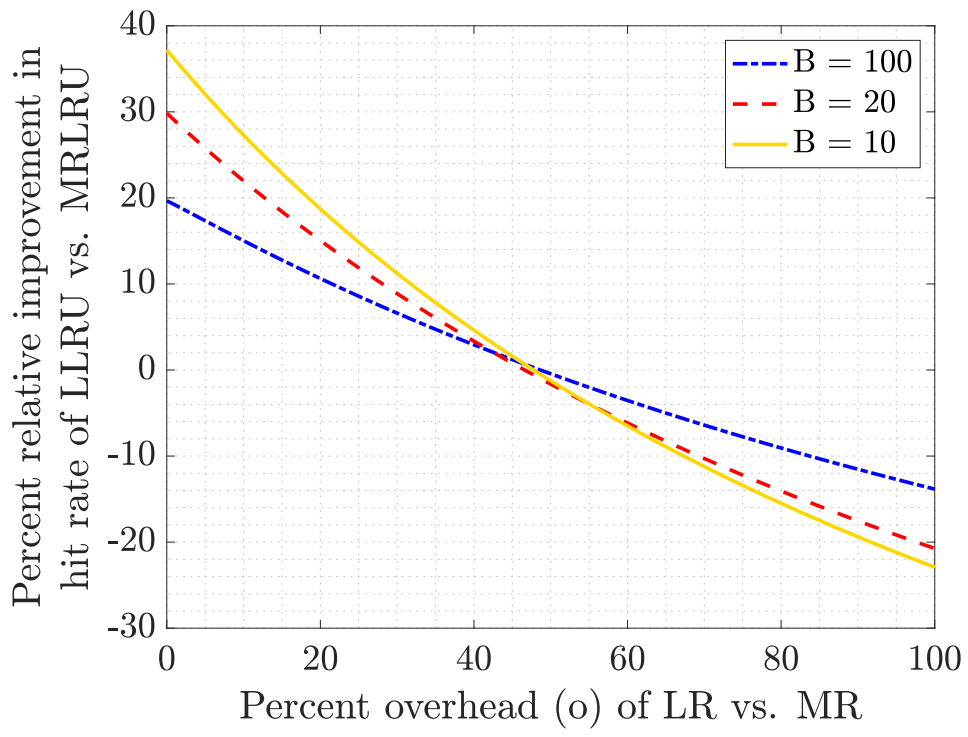

(a) Comparison of hit rate of LLRU to MRLRU for $\alpha = 0.5$.

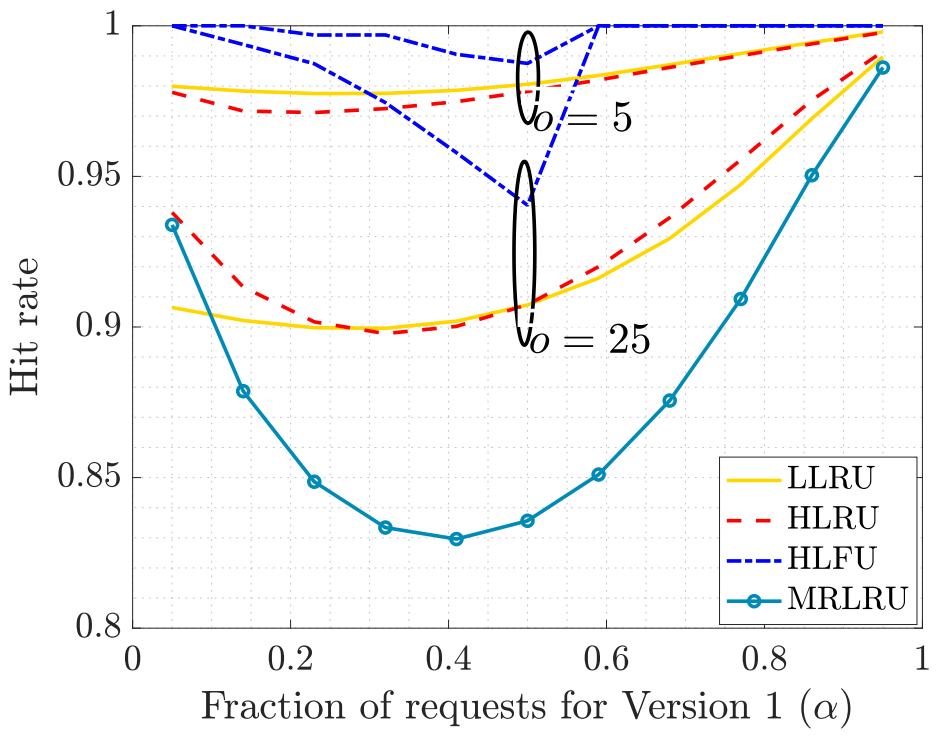

(b) Comparison of hit rate of different caching policies with $B = 100$.

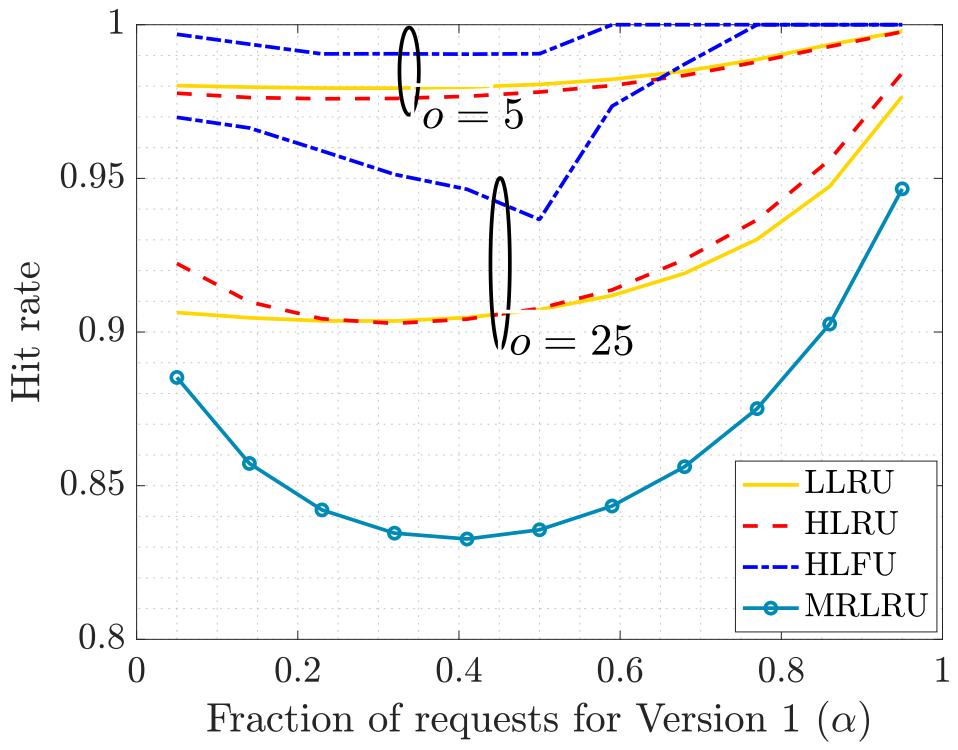

(c) Comparison of hit rate of different caching policies when fraction of requests for versions of odd numbered data objects is $(\alpha, 1 - \alpha)$ and $(0.5, 0.5)$ for the rest with $B = 100$.

Fig. 3. Performance comparison of LR vs. MR for $\beta = 0.5$ as a function of percent overhead and fraction of requests.

We show the results for two different overhead values of 5 and 25. These overhead values represent the extremes for SVC vs. AVC overhead, see [8]. Additionally, we remind the reader of the HLRU and HLFU policies, see Section 2.10, that are capable of adapting the optimal representation for each data object. The rationale behind these approaches is to minimize the storage space occupied by the data object. Initially, the data object is stored in MR, given that $s_{\text{MR}}(d, i) < s_{\text{LR}}(d, i)$ for $i$ equal to 1 or 2. However, if there is an additional request for the other version, the data object is then stored in LR, considering that $s_{\text{LR}}(d, 2) < s_{\text{MR}}(d, 1) + s_{\text{MR}}(d, 2)$.

We note that for $o = 25$, MRLRU performs better or comparable to LLRU when the fraction of requests for different versions is skewed, though this is not necessarily true for lower overhead values for example $o = 5$. Additionally, the hybrid variant of LRU, HLRU, designed to minimize the space occupied by each data object, consistently performs either as well as or better than LLRU for the presented overhead values. We perform a similar study for another scenario where we fix the fraction of requests for Version 1 of even-numbered data objects at 0.5 and vary the fraction of requests for Version 1 of odd-numbered data objects using the parameter $\alpha$, i.e., the fraction of request is $\alpha$ for Version 1 and $1 - \alpha$ for Version 2. The results are plotted in Figure 3c, and once again, the HLRU outperforms or matches LLRU. At last, we draw reader's attention to static Greedy Hybrid, HLFU, which uses the knowledge of popularities of data objects and versions to determine what to cache. This policy consistently outperforms all the other policies irrespective of the overhead values.

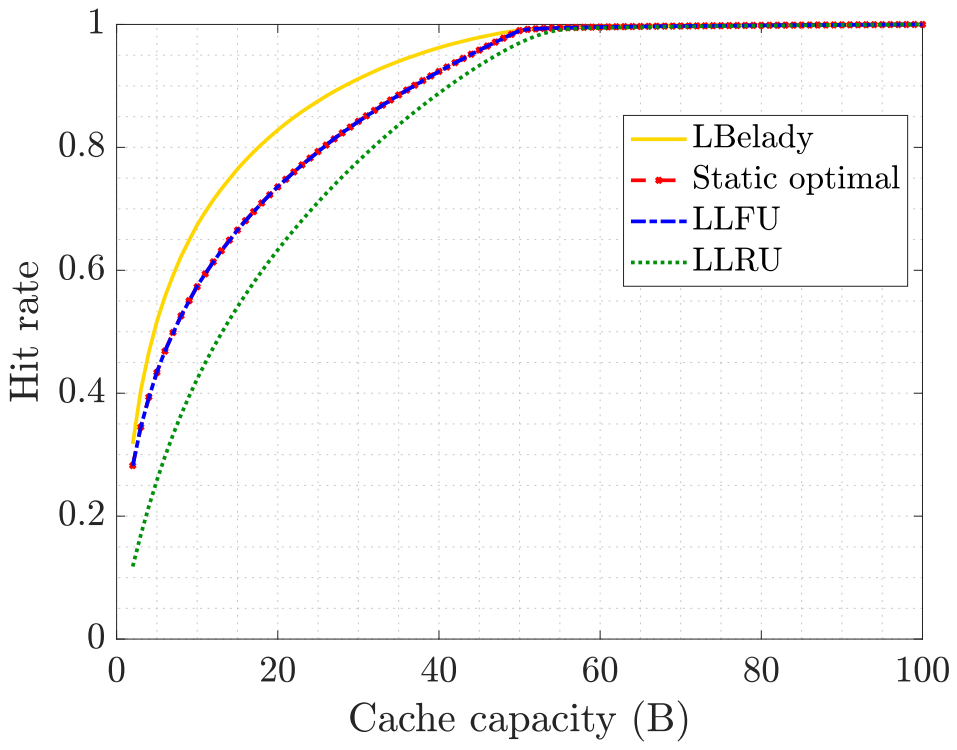

Fig. 4. Hit rate against cache capacity under Layered caching policies for $(\alpha, \rho) = (0.99, 0.5)$.

In summary, we note substantial performance benefits favoring layered representations over multiple representations, especially for reasonable percent overhead ($o$ less than 25). However, for $o = 25$, MRLRU may outperform LLRU if the request distribution is skewed, highlighting the need for policies that can dynamically adapt and store the optimal representation for each data object. Next, we study the different layered caching policies.

## 3.3 Study of different layered caching policies

We begin with a study to compare the performance of different layered caching policies with two vs. only one version under layered representation for each data object. For this, we discretely vary the request rate for LR 1 for a fixed size of Version 1 and 2. Next, we study the performance comparison of LLRU policy for two vs. one version under layered representation as a function of request rate for LR 1 again for a fixed size of Version 1 and 2.

**Setting.** We have a caching system with $D = 100$ data objects. The request probability for data objects follows a Zipf distribution with parameter 0.8. As before $\alpha = q(d, 1)/(q(d, 1) + q(d, 2))$ denotes the fraction of requests for LR 1 of data object $d$ when each data object has two versions. Thus, $1 - \alpha$ is the fraction of requests for LR 2 or requests for both layers. For each data object, $\alpha = 0$ and $\alpha = 1$ correspond to all requests for both layers and only the first layer, respectively. Let $\rho = \delta(d, 1)$ denote the size of Layer 1 for data object $d$, and the total size of each data object is 1, making the size of Layer 2 equal to $1 - \rho$. The request arrival process follows a Poisson distribution with a total request rate equal to 1.

*3.3.1 Layered caching policies.* **Results discussion.** Figure 4 depicts the hit rate for different layered caching policies and the optimal hit rate as a solution of the static optimal policy. We observe that the policy with knowledge of future arrivals, LBelady performs the best, following

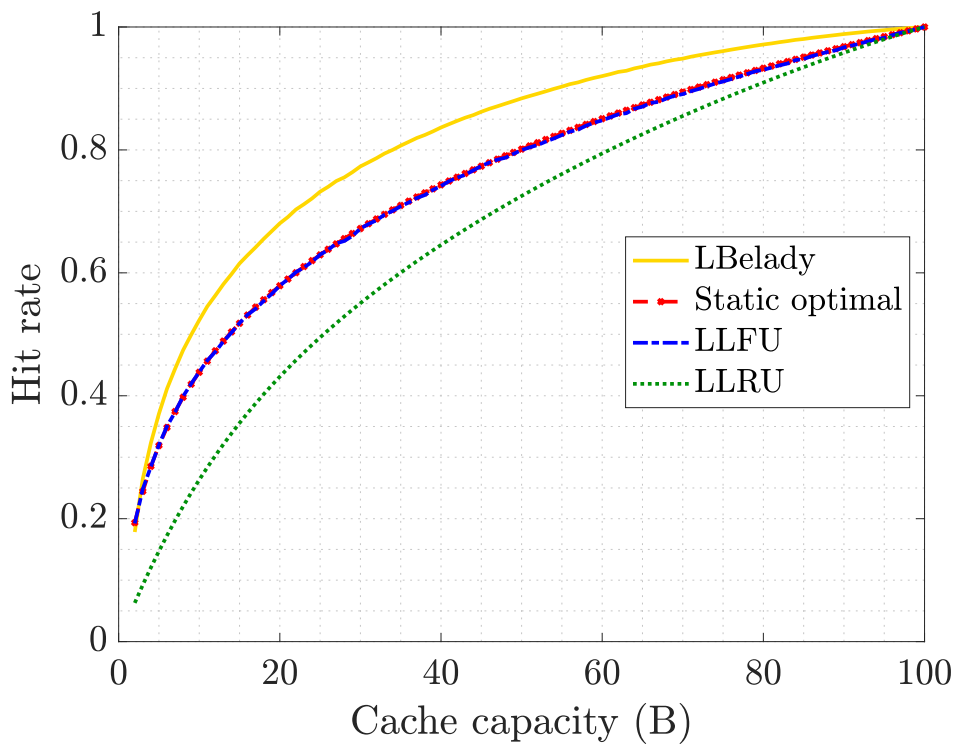

(a) Performance of different caching policies with 1 version under LR for each data object.

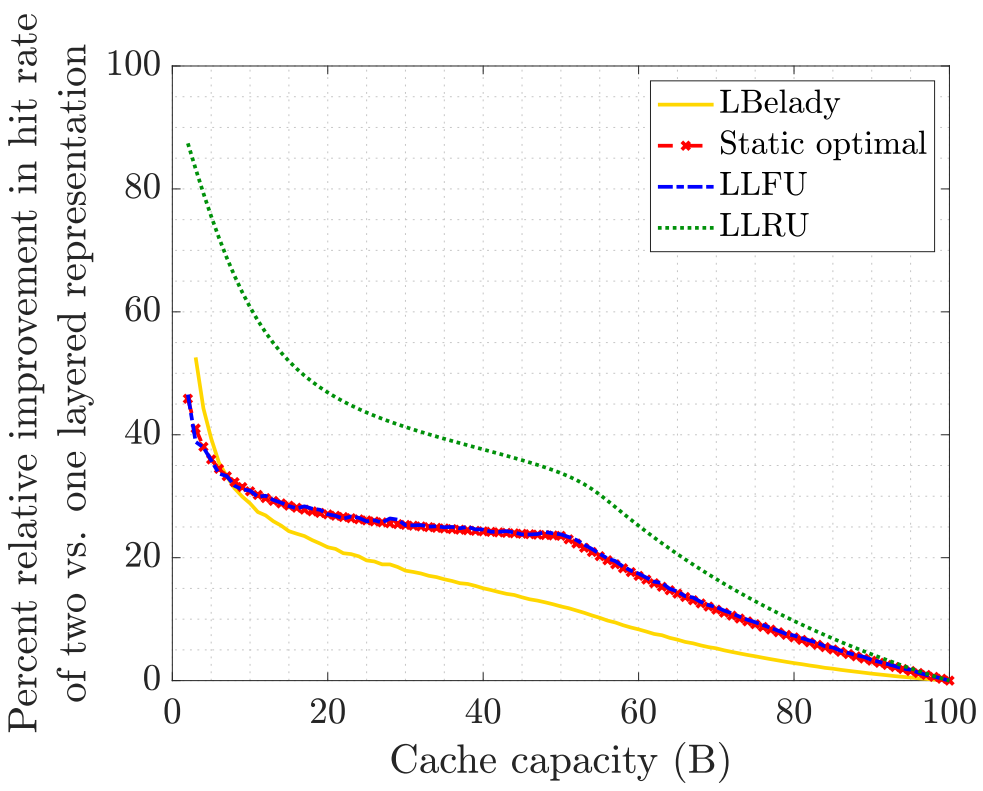

(b) Performance comparison for $(\alpha, \rho) = (0.99, 0.5)$.

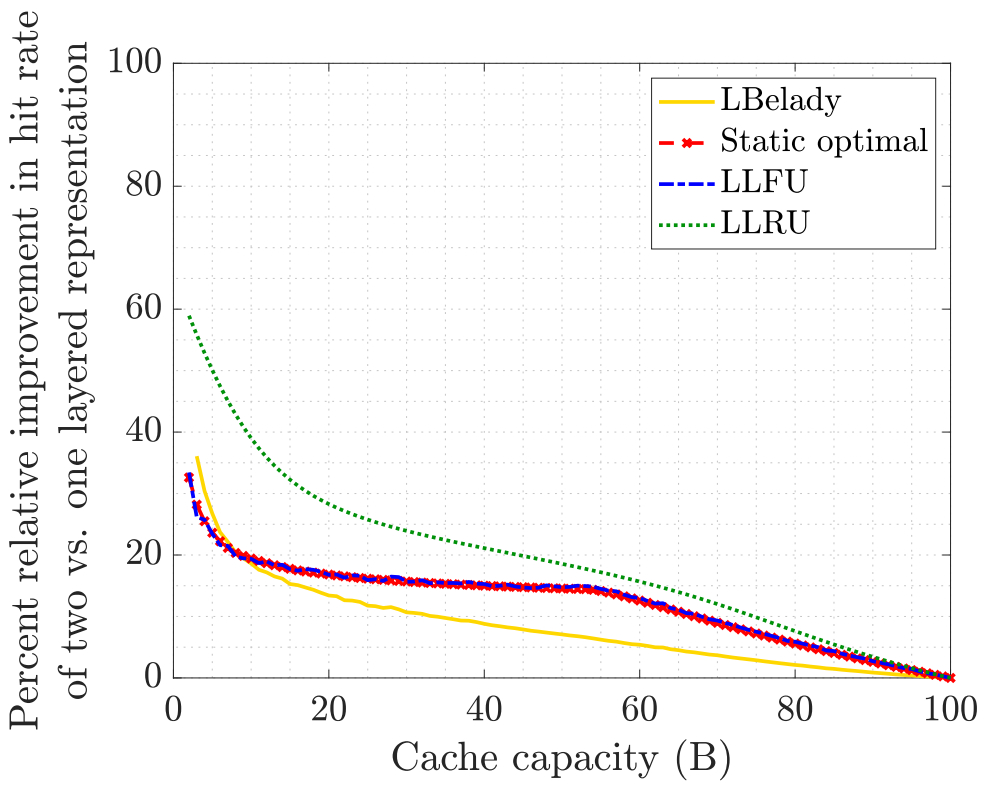

(c) Performance comparison for $(\alpha, \rho) = (0.9, 0.5)$.

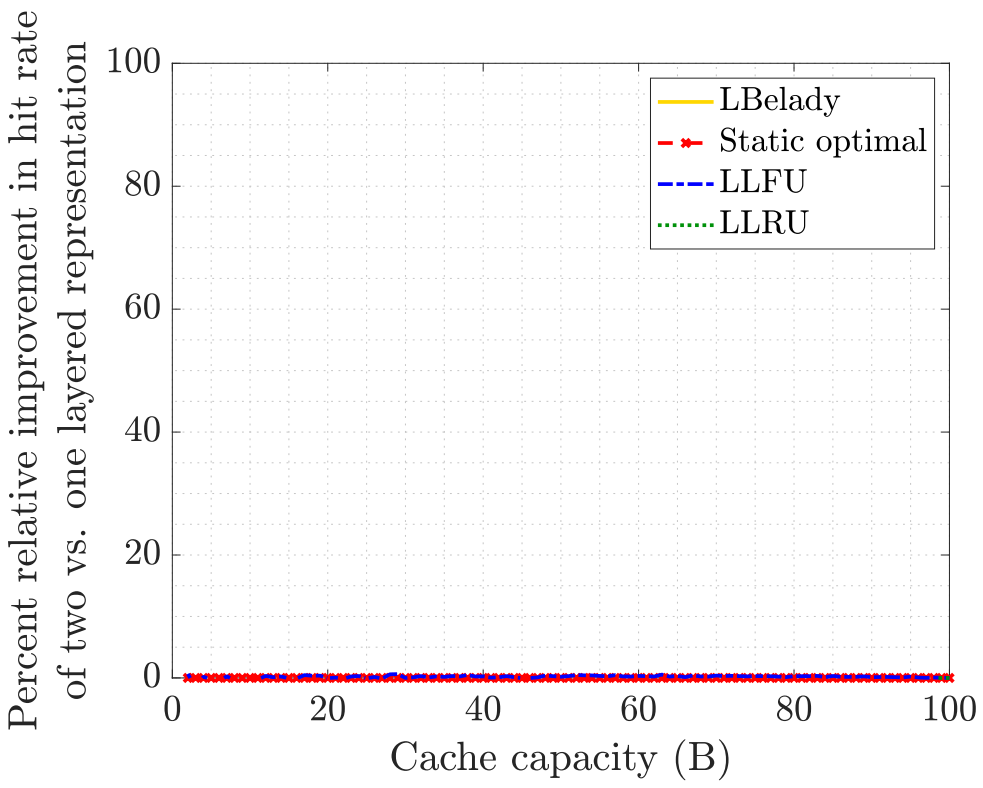

(d) Performance comparison for $(\alpha, \rho) = (0.5, 0.5)$.

Fig. 5. Performance comparison of two vs. one version under layered representation against cache capacity.

that both LLFU and Static optimal have similar performance. Thus, LLFU, which keeps track of the number of arrivals for each data object and version, is the optimal policy among the class of policies without the knowledge of future arrivals. Finally, the LLRU policy, which does not require knowledge of the arrival process nor keep track of the number of arrivals for each data object and version has comparable performance.

*3.3.2 Comparison of hit rate for two vs. one version under layered representation for different caching policies: discrete values for fraction of requests for version 1.* **Results discussion.** As a baseline, we first show the hit rate under different layered caching policies when each data object consists of only 1 version in Figure 5a. We then plot the percent relative improvement in hit rate of two vs. one version under layered representation with different layered caching policies for different values of $\alpha$ and $\rho$ in Figure 5. As $\alpha$ increases for fixed $\rho$, we observe an improvement in relative performance for all policies at a given cache capacity. Additionally, for cache capacity equal to 100 the percent relative improvement is 0 because all policies achieve the best possible hit rate.

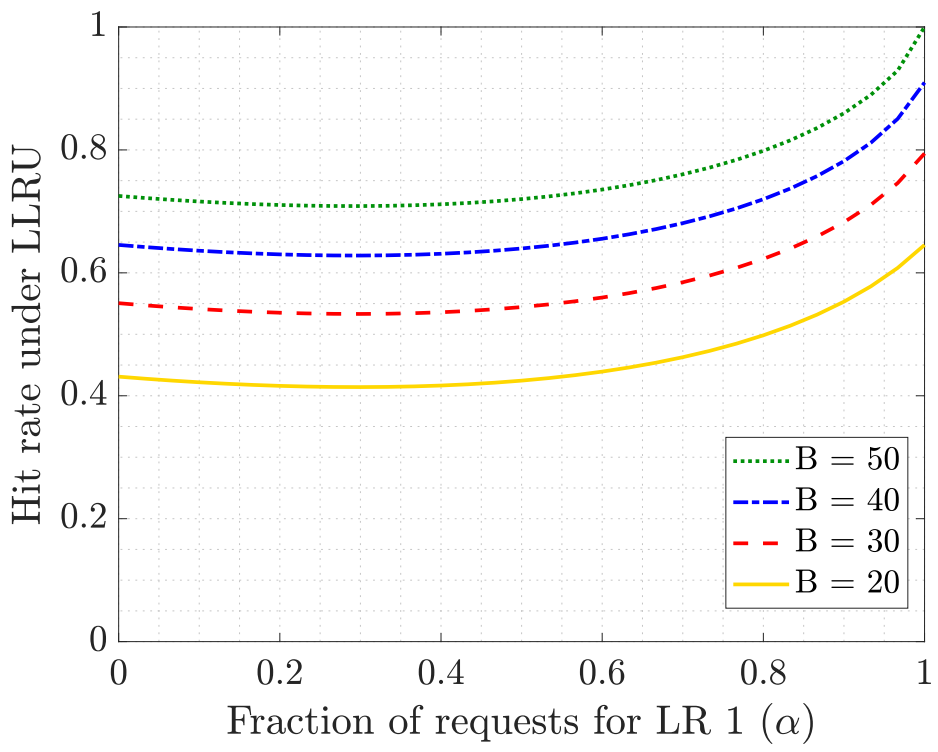

(a) Hit rate under LLRU with two versions under layered representation for each data object.

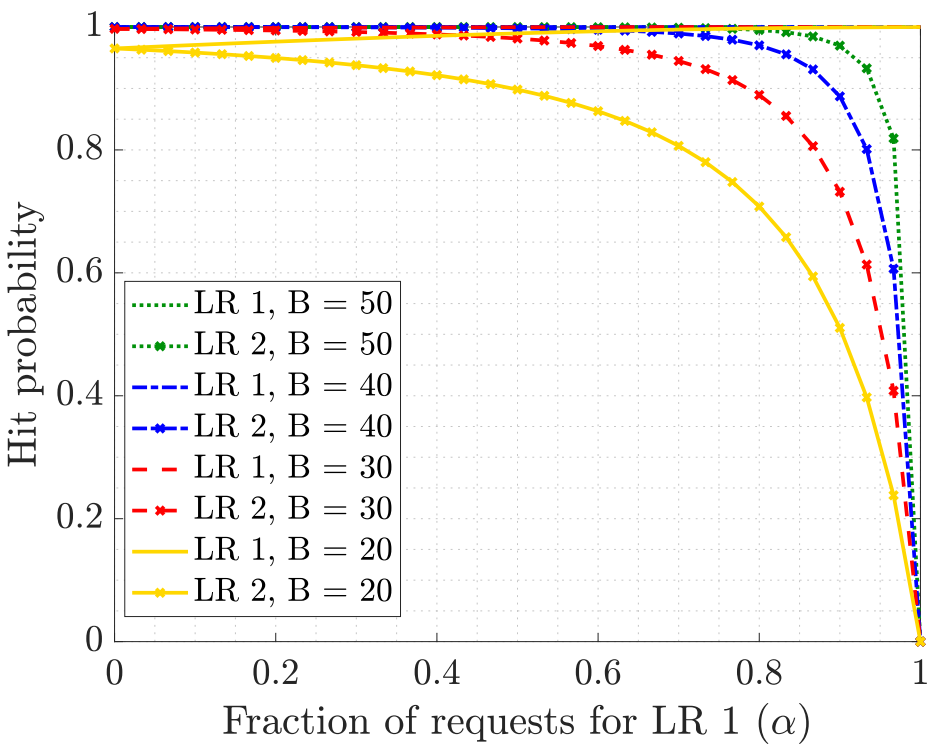

(b) Hit probability for Data Object 1.

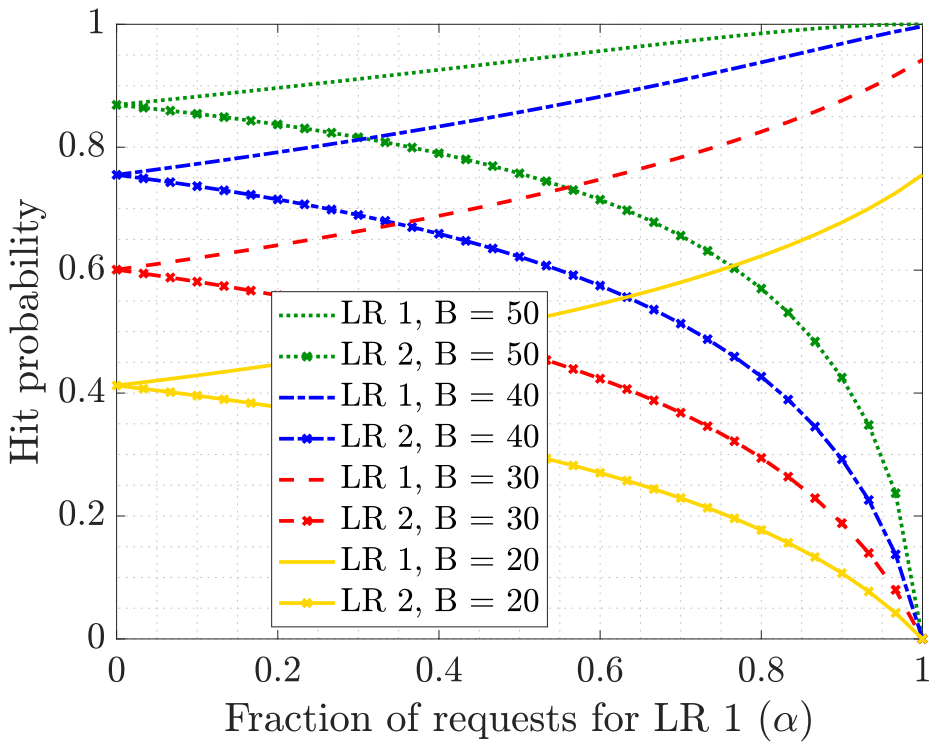

(c) Hit probability for Data Object 10.

Fig. 6. Performance of LLRU with two versions under layered representation for each data object against fraction of requests for LR 1 and $\rho = 0.5$.

*3.3.3 Comparison of hit rate for two vs. one version under layered representation under LLRU: varying fraction of requests for Version 1.* **Results discussion.** Figure 6a shows the hit rate under the LLRU caching policy for different cache capacities as a function of the fraction of requests for LR 1, $\alpha$. We observe a non-monotonic behavior for the hit rate for all cache capacities. This is explained through Figures 6b and 6c, where we show the hit probability for both versions of data object 1 and 10, respectively, for different cache capacities. In both figures, the hit probability for LR 1 increases as the fraction of requests for Version 1 increases from left to right, while decreasing for LR 2. Since the hit rate is a convex combination of hit probabilities for LR 1 and LR 2, we observe the non-monotonic behavior in Figure 6a. Thus, for a fixed size of Version 1, the performance is non-monotonic in the fraction of requests for LR 1.

### 3.4 Impact of layers' sizes and popularity on performance

In this section, we study the impact of layer size and popularity of layers on the hit rate. More specifically, we will fix the size of layers and offer guidance on how to set the popularity of versions and thus layers that is optimal. Similarly, for a fixed popularity of versions, we address the optimal setting of the size of each layer. We will do this first for the case where data objects have 2 versions and then 3 versions.

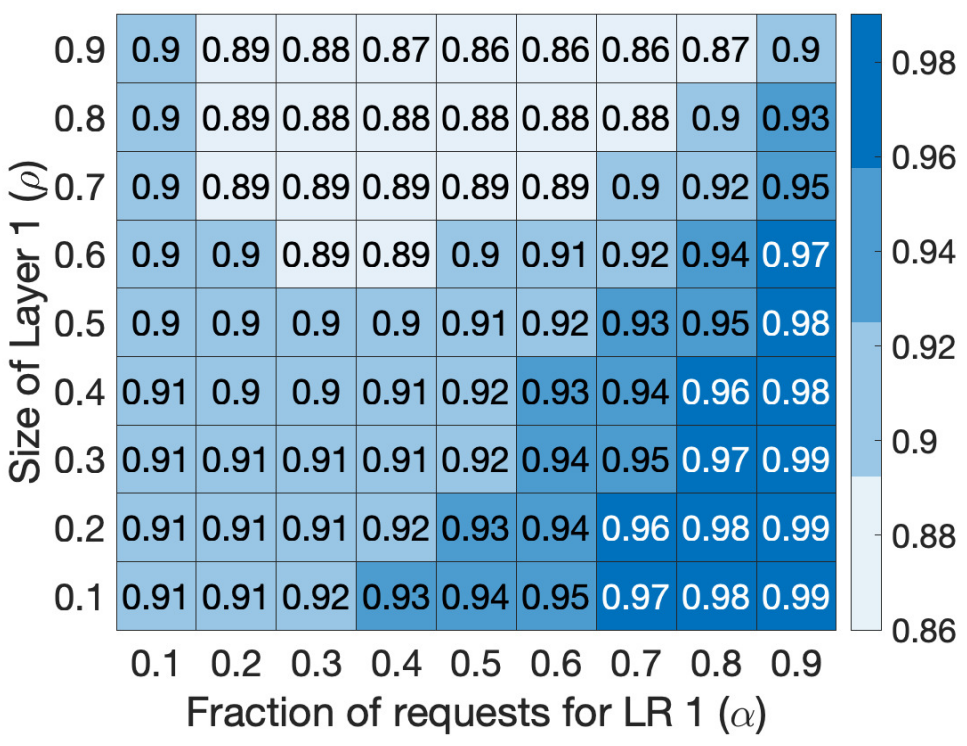

Fig. 7. Performance of LLRU for different size and popularity for a cache capacity of 20.

**Setting.** For this section, we have a caching system with $D = 100$ data objects. The request probability for data objects follows a Zipf distribution with parameter 0.8. For the case of 2 versions, we remind the reader about $\alpha$, which denotes the fraction of requests for Version 1 of data object $d$ and $\rho = \delta(d, 1)$ denotes the size of Layer 1 of data object $d$. With 3 versions for each data object, we let $\zeta = q(d, 1)/(q(d, 1) + q(d, 2) + q(d, 3))$, and $\eta = q(d, 2)/(q(d, 1) + q(d, 2) + q(d, 3))$ denote the fraction of requests for Version 1 and 2 respectively. Thus, the fraction of requests for Version 3 is $1 - \zeta - \eta$. We use $\rho = \delta(d, 1)$ and $\kappa = \delta(d, 2)$ to denote the size of Layer 1 and Layer 2, making the size of Layer 3 equal to $1 - \rho - \kappa$.

*3.4.1 How to set the size and popularity when each data object has 2 version?* **Results discussion.** We show the performance of cache under LLRU caching policy in Figure 7 for different values of fraction of requests for Version 1 and size of Layer 1. We observe that for a fixed fraction of requests for LR 1, as the size of Layer 1 decreases, the performance improvement is monotonically increasing. Also, as already observed in the previous section, the same is not true for the fixed size of Layer 1 and the increasing fraction of requests for LR 1. Furthermore, if both the size and fraction of requests vary simultaneously, possibly along a diagonal, the hit rate does not follow a monotonic pattern. The last observation is significant improvements occur with an increasing fraction of requests for LR 1 and a decreasing size of Layer 1.

*3.4.2 How to set the size and popularity when each data object has 3 versions?* **Results discussion.** We show the performance of cache under LLRU caching policy in Figure 8 for different values of fraction of requests for LR 1 and LR 2 under different fixed sizes of layers. We limit ourselves to a scenario where the fraction of requests for any version is at least 0.1 and thus for infeasible pairs of $(\zeta, \eta)$ we set the hit rate value to 0. In Figure 8a, we observe that the maximum hit rate is observed for $(\zeta, \eta) = (0.8, 0.1)$, i.e., if most of the requests are for LR 1, which also has a small size, one observes the maximum hit rate. Similarly, in Figure 8b we observe the maximum hit rate for $(\zeta, \eta) = (0.8, 0.1)$ even though the size of layer 1 is the maximum of all. This is a result of the condition that the presence of a higher layer implies all layers lower than that must also be present in the cache. In addition, we observe that for a fixed value of $\zeta$ and increasing $\eta$, the hit rate is not monotonic. Thus, a naive approach to selecting the popularity might not be optimal. We show

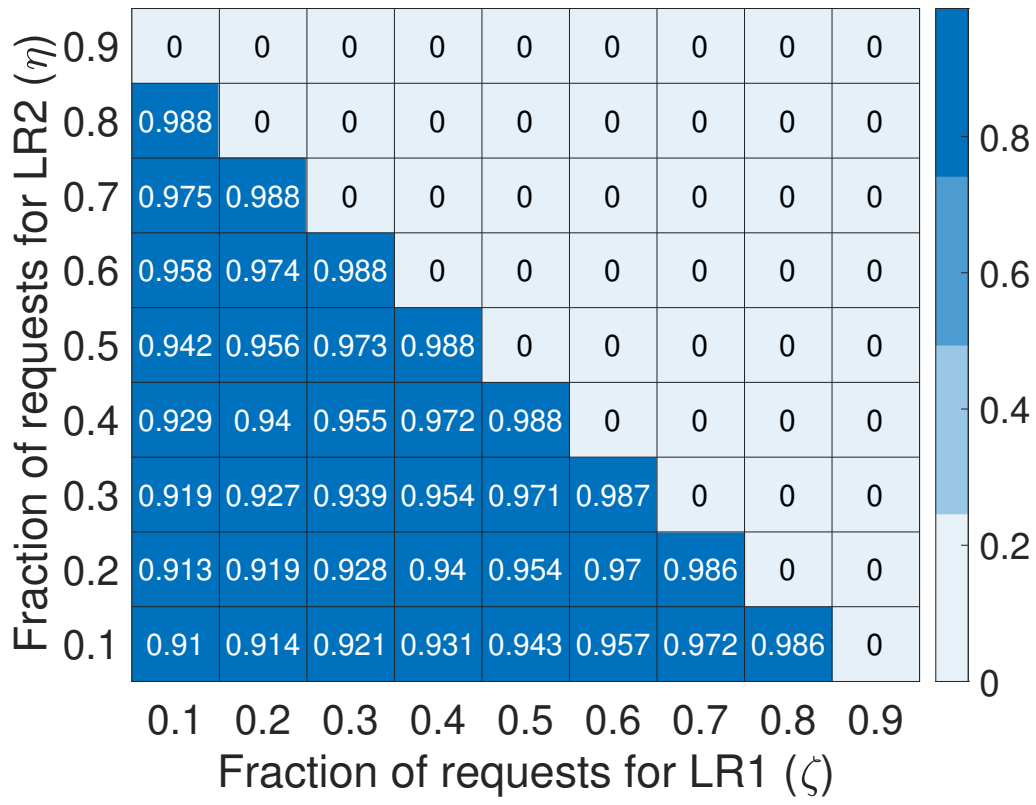

(a) Performance for $(\rho, \kappa) = (0.1, 0.1)$.

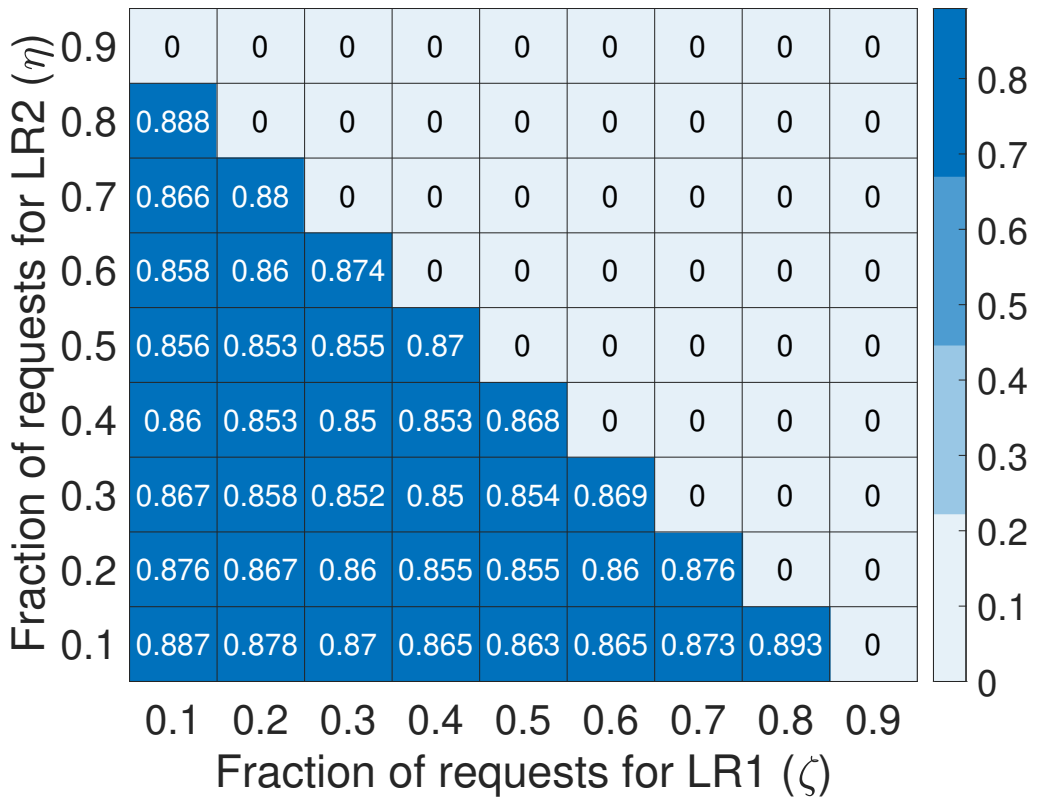

(b) Performance for $(\rho, \kappa) = (0.8, 0.1)$.

Fig. 8. Performance of LLRU for different popularities of versions when there 3 versions for $B = 80$.

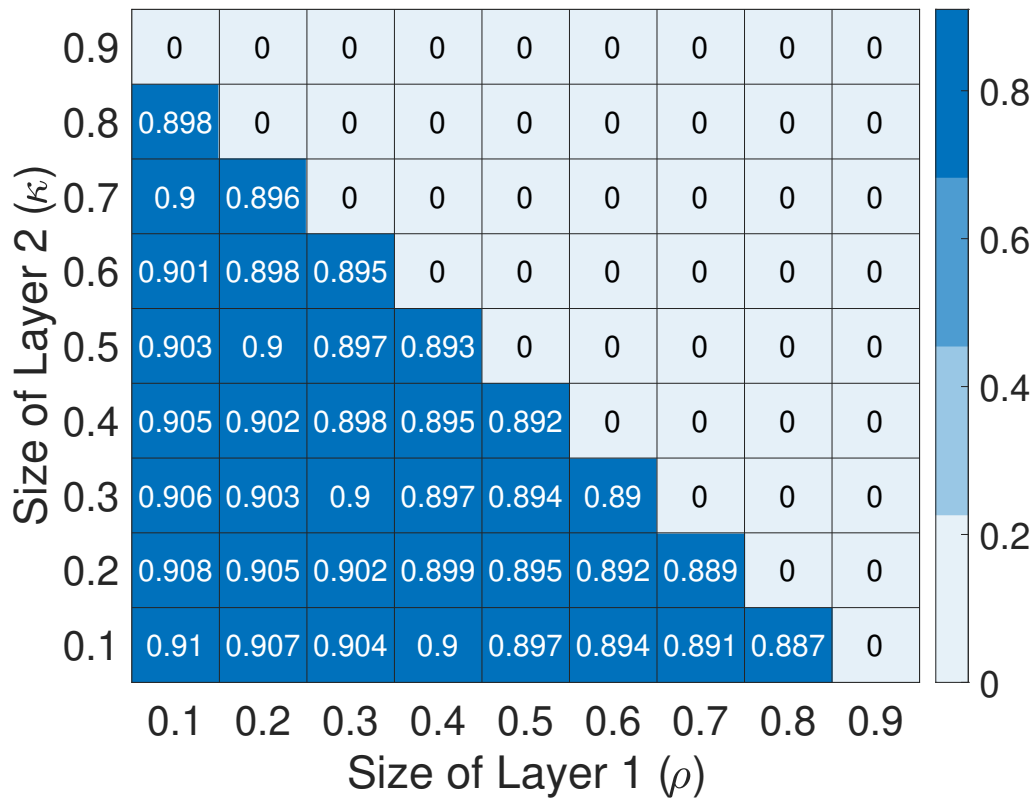

(a) Performance for $(\zeta, \eta) = (0.1, 0.1)$.

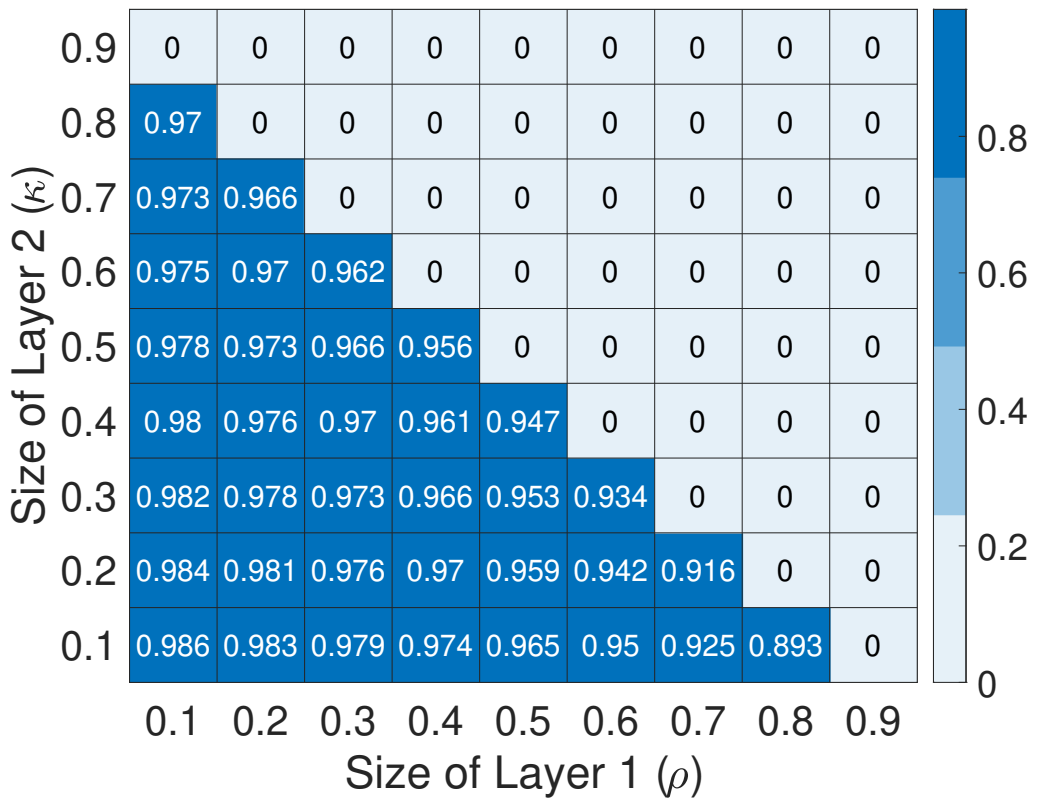

(b) Performance for $(\zeta, \eta) = (0.8, 0.1)$.

Fig. 9. Performance of LLRU for different popularities of versions when there 3 versions for $B = 80$.

similar results for the case when popularity is fixed, and we need to select the optimal size of layers in Figure 9.

### 3.5 Is it beneficial to increase the number of versions for a data object?

**Setting.** We manage 100 data objects, and the request probability for each data object follows a Zipf distribution with parameter 0.8. We scale the number of versions as $V$ and correspondingly vary the request probability for data object $d$'s $v$th version as $q^{(V)}(d, v) = \frac{(V-v+1)^m}{\sum_{i=1}^{V}(V-i+1)^m}$ while size varies as $s_{\text{LR}}^{(V)}(d, v) = \sum_{l=1}^{v} \delta^{(V)}(d, l)$ where $\delta^{(V)}(d, l) = \frac{(l)^n}{\sum_{i=1}^{V}(i)^n}$. We plot the request probability, $p^{(V)}(d, l) = \sum_{v=l}^{V} q^{(V)}(d, v)$, for the first three layers and size, $\delta^{(V)}(d, l)$, as a function of number of versions in Figure 10a and Figure 10b, respectively.

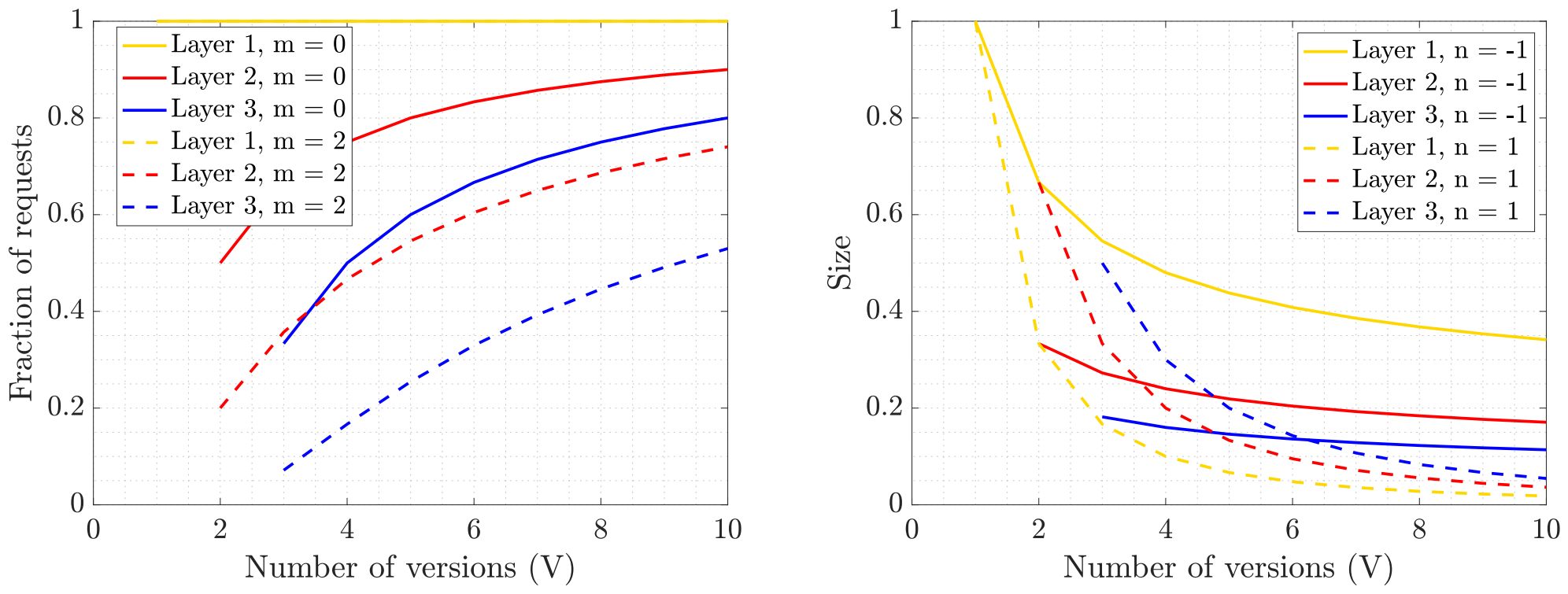

(a) Fraction of requests for layers.

(b) Size of layers.

Fig. 10. Popularity and size characterization for different values of $m$ and $n$ as a function of number of versions.

**Results discussion.** We conduct a performance comparison of LLRU with V vs. 1 versions under layered representation in Figure 11. In Figures 11a, 11b, 11c, and 11d we show the hit rate under LLRU against the number of versions for different values of $m$ and $n$. We empirically observe that both the popularity and size of layers need to increase/decrease at a certain rate to see benefits in terms of the hit rate. In Figure 11a, the hit rate is monotonically decreasing in the number of versions whereas by increasing the value of $m$ for same $n$, we see a non-monotonic behavior, see Figure 11b. This points to the subtle ways in which the overall hit rate depends on the number of versions, popularity, and size characterization. In addition, we observed that the hit rate is monotonic in the number of versions for all values of $m > 0$ and $n \geq 0$.

## 4 CONCLUSION

The efficient management of the large amounts of data required by emerging delay-constrained applications, e.g., multiplayer VR gaming and NN-based inference, will require judicious use of caching which exploit, when appropriate, hierarchies of data object representations that enable tradeoffs between data object's size and quality. To address this, in this paper, we have studied caching policies optimized for data objects with multiple versions and layered representations. Based on numerical analysis and simulation, the benefits of LR are substantial even if in some settings such hierarchical representations incur additional overheads. To make the most of such representations it is critical to understand the impact that the incremental size of layers and the level of demand for different versions will play. This paper explores these impacts and suggests when, for example, additional layers may be of value, and when they may in be counterproductive, towards enhancing performance.

## ACKNOWLEDGEMENTS

This work was supported by National Science Foundation CNS- 2212202 Award.

## APPENDIX

In this section we provide the proof our theorem for hit probability of a data object and layer.

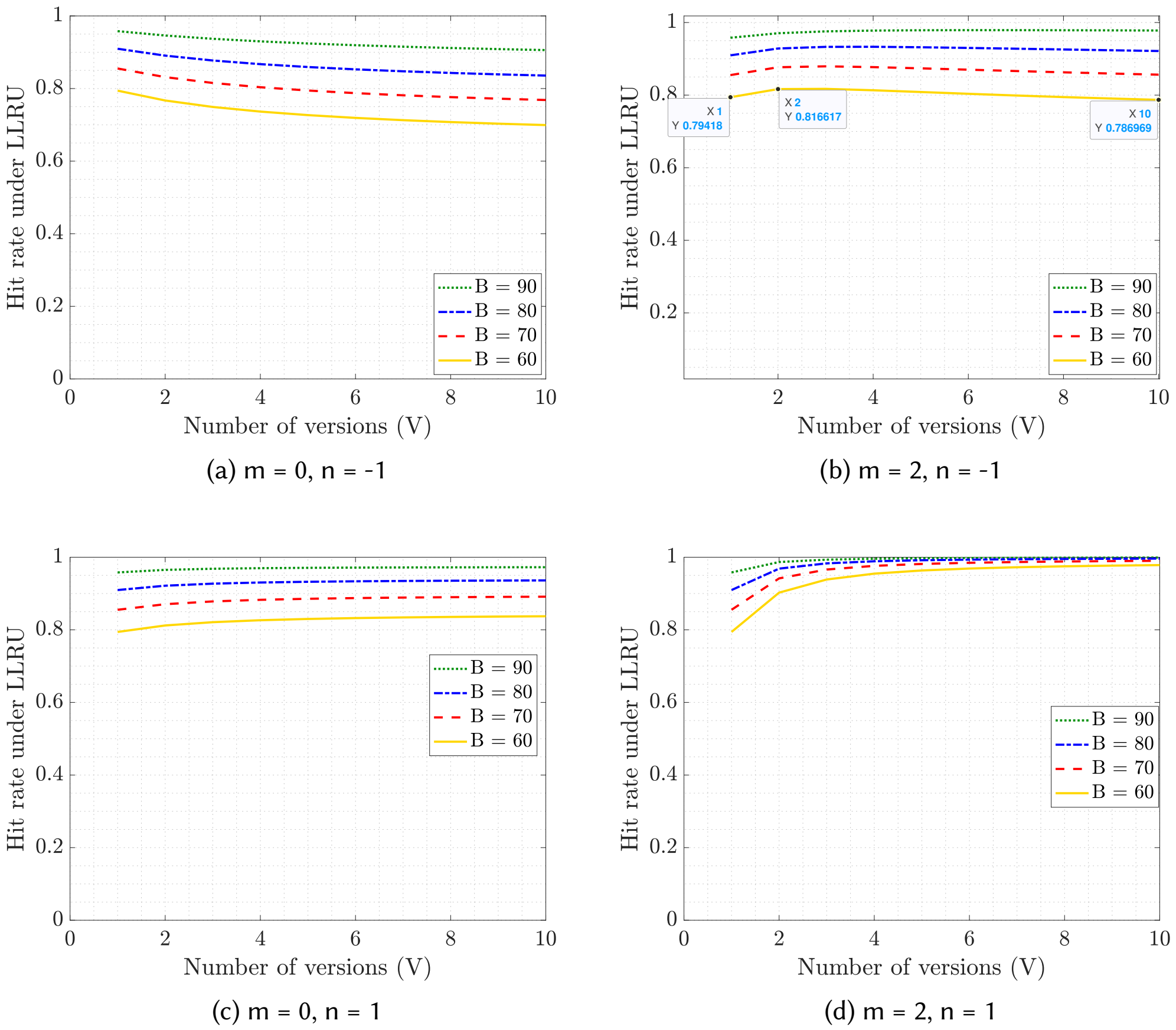

Fig. 11. Comparison of LLRU with V vs. 1 version under layered representation for request probability of version $v$ of data object $d$ given by $q^{(V)}(d,v) = \frac{(V-v+1)^m}{\sum_{i=1}^{V}(V-i+1)^m}$ and size of $l$th layer by $\delta^{(V)}(d,l) = \frac{(l)^n}{\sum_{i=1}^{V}(i)^n}$.

Lemma 3. *Under the working set memory management in the independent reference model with $D$ data objects and $V$ layers, the variance in the size of working set is bounded above by* $\left(\frac{D\cdot V}{4} + D\cdot V\cdot(V-1)\right)\cdot\left(\delta_{\max}^{(D)}\right)^2$, *where* $\delta_{\max}^{(D)} = \max_{d\in\mathcal{D},l\in\{1,2,\cdots,V\}}\delta^{(D)}(d,l)$

Proof. Let $X^{(D)}(d,l)$ be a random variable which is 1 if data object $d$ and layer $l$ is in the cache at time $t$, and 0 otherwise. The working set at time $t$ is given by:

$$S^{(D)}(t) = \sum_{d=1}^{D}\sum_{l=1}^{V}\delta^{(D)}(d,l)X^{(D)}(d,l), \tag{28}$$

then the variance $\mathrm{V}\left(S^{(D)}(t)\right)$ in the size of working set is

$$\begin{aligned}\mathrm{V}\left(S^{(D)}(t)\right) &= \mathrm{V}\left(\sum_{d=1}^{D}\sum_{l=1}^{V}\delta^{(D)}(d,l)X^{(D)}(d,l)\right)\\ &= \sum_{d=1}^{D}\sum_{l=1}^{V}\left(\delta^{(D)}(d,l)\right)^2 \mathrm{V}\left(X^{(D)}(d,l)\right) + \\ &\quad 2\sum_{1\le i<d\le D}\sum_{l=1}^{V}\left(\delta^{(D)}(i,l)\delta^{(D)}(d,l)\right)\mathrm{Cov}\left(X^{(D)}(i,l),X^{(D)}(d,l)\right) + \\ &\quad 2\sum_{d=1}^{D}\sum_{1\le l<k\le V}\left(\delta^{(D)}(d,l)\delta^{(D)}(d,k)\right)\mathrm{Cov}\left(X^{(D)}(d,l),X^{(D)}(d,k)\right)\end{aligned}$$

where Cov is the covariance. Since

$$\begin{aligned}\mathrm{V}\left(X^{(D)}(d,l)\right) &\le 1/4,\\ \mathrm{Cov}\left(X^{(D)}(i,l),X^{(D)}(d,l)\right) &\le 0, i\neq d\\ \mathrm{Cov}\left(X^{(D)}(d,l),X^{(D)}(d,k)\right) &\le 1, l\neq k,\end{aligned}$$

We find that

$$\mathrm{V}\left(S^{(D)}(t)\right) \le \left(\frac{D\cdot V}{4} + D\cdot V\cdot(V-1)\right)\cdot\left(\delta^{(D)}_{\max}\right)^2 \tag{29}$$

□

## 4.1 Proof of 1

We first show that

$$\lim_{D\to\infty}\mathbb{E}\left[\frac{S^{(D)}(D\tau)}{D}\right] = \int_0^1\sum_{l=1}^{V}\Delta(x,l)dx - \int_0^1\sum_{l=1}^{V}\Delta(x,l)e^{-\tau F'(x)\sum_{v=l}^{V} g(v;x)}dx \tag{30}$$

where

$$\mathbb{E}\left[\frac{S^{(D)}(D\tau)}{D}\right] = \frac{1}{D}\sum_{d=1}^{D}\sum_{l=1}^{V}\delta^{(D)}(d,l)\left(1-(1-p^{(D)}(d,l))^{(D\tau-1)}\right) \tag{31}$$

By the Mean Value Theorem,

$$q^{(D)}(d,v) = (F'(\psi(d))/D)\cdot g(v;d/D) \tag{32}$$

for some $\psi(d)$ with $((d-1)/D) \le \psi(d) \le (d/D)$ and $p^{(D)}(d,l) = (F'(\psi(d))/D)\sum_{v=l}^{V} q^{(D)}(d,v)$

We now use Lemma 10 from [7], which states that for each closed bounded set $C$,

$$(1-(c/n))^{\tau_0 n} \longrightarrow e^{-\tau_0 c} \quad \text{as } n\longrightarrow\infty, \text{uniformly over all } c \text{ in } C. \tag{33}$$

The above is just using point wise limits. Thus, for $D \gg 1$ if

$$\left|\left(1-\frac{F'(\psi(d))}{D}\sum_{v=l}^{V} q^{(D)}(d,v)\right)^{D\tau_0} - e^{-\tau_0 F'(\psi(d))\sum_{v=l}^{V} q^{(D)}(d,v)}\right| < \epsilon \tag{34}$$

then one can easily show the following using the same arguments as from [7]

$$\left|\frac{1}{D}\sum_{d=1}^{D}\sum_{l=1}^{V}\delta^{(D)}(d,l)\left((1-p^{(D)}(d,l))^{(D\tau_0-1)}-e^{-\tau_0 F'(\psi(d))\sum_{v=l}^{V}q^{(D)}(d,v)}\right)\right|<\epsilon \tag{35}$$

Now by the definition of Riemann Integral,

$$\frac{1}{D}\sum_{d=1}^{D}\sum_{l=1}^{V}\delta^{(D)}(d,l)\left(1-e^{-\tau_0 F'(\psi(d))\sum_{v=l}^{V}q^{(D)}(d,v)}\right) \tag{36}$$

is an approximation to the following integral

$$\int_0^1\sum_{l=1}^{V}\Delta(x,l)dx-\int_0^1\sum_{l=1}^{V}\Delta(x,l)e^{-\tau F'(x)\sum_{v=l}^{V}g(v;x)}dx \tag{37}$$

where Riemann integrable $\Delta$ satisfies $\Delta(d/D,l)=\delta^{(D)}(d,l)$ for all $D,d$ and $l$. The absolute error between Eq. 36 and Eq. 37 can be made smaller than $\epsilon$ for sufficiently large $D$. Thus, we show the result in Eq. 30.

Additionally, using Lemma 3, we obtain the following

$$\lim_{D\to\infty}\mathrm{V}\left(\frac{S^{(D)}(D\tau)}{D}\right)\to 0. \tag{38}$$

Let $\tau^*$ denote a unique solution to

$$b=\lim_{D\to\infty}\mathbb{E}\left[\frac{S^{(D)}(D\tau)}{D}\right]=\int_0^1\sum_{l=1}^{V}\Delta(x,l)dx-\int_0^1\sum_{l=1}^{V}\Delta(x,l)e^{-\tau F'(x)\sum_{v=l}^{V}g(v;x)}dx \tag{39}$$

For finite $D\gg 1$, this equation is approximated by

$$B=\mathbb{E}\left[S^{(D,V)}(t)\right]=\sum_{d=1}^{D}\sum_{l=1}^{V}\delta^{(D)}(d,l)\left(1-(1-p^{(D)}(d,l))^{(D\tau-1)}\right) \tag{40}$$

with $t^*=D\tau^*$ as the unique solution for the above equation when $B=Db$.

Note that as $D\longrightarrow\infty$,

$$\frac{S^{(D)}_{-(d,l)}(D\tau)}{D}\sim\frac{S^{(D)}(D\tau)}{D}$$

So,

$$\begin{aligned}\lim_{D\to\infty}\mathbb{P}\left(S^{(D)}_{-(d,l)}(D\tau)\geq B\right)&=\lim_{D\to\infty}\mathbb{P}\left(S^{(D)}(D\tau)/D\geq b\right)\\&=u(\tau-\tau^*)\end{aligned}$$

By Palm's theorem [2], the stationary LRU miss probability for data object $d$ and layer $l$ is

$$\begin{aligned}1-h^{(D)}(d,l)&=\mathbb{P}\left(S^{(D)}_{-(d,l)}(T_n^{(D)(d,l)})\geq B\right)\\&=\sum_{t=1}^{\infty}\mathbb{P}\left(S^{(D)}_{-(d,l)}(t)\geq B\right)p^{(D)}(d,l)(1-p^{(D)}(d,l))^{t-1}\end{aligned}$$

For $t = D\tau$, $B = Db$, $D \gg 1$, we can obtain the following with $\tau^*$ as the unique solution of Eq. 39

$$
\begin{aligned}
1 - h^{(D)}(d,l) &= \sum_{\tau=1/(D)}^{\infty} u(\tau - \tau^*) p^{(D)}(d,l)(1 - p^{(D)}(d,l))^{D\tau - 1} \\
&= (1 - p^{(D)}(d,l))^{D\tau^* - 1} \\
&= (1 - p^{(D)}(d,l))^{t^* - 1}
\end{aligned}
$$

for all data objects $d$ and layer $l$. As $D \longrightarrow \infty$, using Lemma 10 from [7] or point wise limits for right hand side, we obtain

$$1 - h(d,l) = e^{-\tau^* F'(d) \sum_{v=l}^{V} g(v;d)}. \tag{41}$$